\documentclass[
  aps, pra,
 amsmath,amssymb,
 11pt,
 final,
tightenlines,
 twoside,
 twocolumn,
 nofloats,
nofootinbib,
 superscriptaddress,
showkeys,
showkeywords,
 ]
{revtex4}

\usepackage[T2A]{fontenc}
\usepackage[utf8x]{inputenc}
\usepackage[english]{babel}
\usepackage{graphicx}% Include figure files
\usepackage{dcolumn}% Align table columns on decimal point
\usepackage{bm}% bold math
\usepackage{longtable}

\usepackage{subfigure}

\input{maik.rty}

\setcitestyle{authoryear,round,aysep={,},citesep={;}}
\def\saoname{Special Astrophysical Observatory,  Russian Academy of Sciences,
Nizhnii Arkhyz, 369167 Russia}

\def\squareforqed{\hbox{\rlap{$\sqcap$}$\sqcup$}}

\def\sq{\ifmmode\squareforqed\else{\unskip\nobreak\hfil
\penalty50\hskip1em\null\nobreak\hfil\squareforqed
\parfillskip=0pt\finalhyphendemerits=0\endgraf}\fi}

\def\utw{\smash{\rlap{\lower5pt\hbox{$\sim$}}}}

\def\udtw{\smash{\rlap{\lower6pt\hbox{$\approx$}}}}

\def\fd{\hbox{$\,.\!\!^{\rm d}$}}

\def\fm{\hbox{$\,.\!\!^{\rm m}$}}

\def\diameter{{\ifmmode\mathchoice
{\ooalign{\hfil\hbox{$\displaystyle/$}\hfil\crcr
{\hbox{$\displaystyle\mathchar"20D$}}}}
{\ooalign{\hfil\hbox{$\textstyle/$}\hfil\crcr
{\hbox{$\textstyle\mathchar"20D$}}}}
{\ooalign{\hfil\hbox{$\scriptstyle/$}\hfil\crcr
{\hbox{$\scriptstyle\mathchar"20D$}}}}
{\ooalign{\hfil\hbox{$\scriptscriptstyle/$}\hfil\crcr
{\hbox{$\scriptscriptstyle\mathchar"20D$}}}}
\else{\ooalign{\hfil/\hfil\crcr\mathhexbox20D}}%
\fi}}

\newcommand{\ab}{Astrophysical Bulletin }
\newcommand{\mnras}{Monthly Notices Royal Astron. Soc. }
\begin{document}

\selectlanguage{english}


\title{Magnetic Fields of New CP Stars Discovered with Kepler Mission Data}

\author{\firstname{I. A.}~\surname{Yakunin}} \email{elias@sao.ru}
\affiliation{St.~Petersburg University, St.~Petersburg, 199034 Russia}
\affiliation{Special Astrophysical Observatory,  Russian Academy of Sciences,
Nizhnii Arkhyz, 369167 Russia}

\author{\firstname{E. A.}~\surname{Semenko}}
 \affiliation{National Astronomical Research Institute of Thailand, Mae Rim, Chiang Mai 50180, Thailand} \affiliation{\saoname}

\author{\firstname{I. I.}~\surname{Romanyuk}}
 \affiliation{Special Astrophysical Observatory,  Russian Academy of Sciences,
Nizhnii Arkhyz, 369167 Russia}

\author{\firstname{A. V.}~\surname{Moiseeva}}
\affiliation{Special Astrophysical Observatory,  Russian Academy of Sciences,
Nizhnii Arkhyz, 369167 Russia}

\author{\firstname{V.~N.}~\surname{Aitov}} \affiliation{Special Astrophysical Observatory,  Russian Academy of Sciences,
Nizhnii Arkhyz, 369167 Russia}

%\received{November 29, 2022} \revised{March 20, 2023} \accepted{March 21, 2023}

\begin{abstract} 
The paper presents the first results of the
ongoing spectropolarimetric monitoring of magnetic fields of
stars, whose chemically peculiar nature has been previously
revealed with the 1-m SAO RAS telescope. We selected the sample
candidates using the photometric data of the Kepler and TESS space
missions. The efficiency of the method of searching for new CP
stars based on photometric light curves has been confirmed. We
present the magnetic field measurements   and
estimate the atmospheric parameters of the objects under study.
\end{abstract}

\maketitle

\section{INTRODUCTION} Chemically peculiar (CP) stars of the upper
part of the Main Sequence (MS) are characterized by the anomalous
abundance of some chemical elements in the photosphere often
unevenly distributed over the stellar surface. Among them, a group
of magnetic chemically peculiar (mCP) stars stands out; it
consists of classical Ap/Bp (CP2) stars and He-weak/rich (CP4)
stars (Preston, 1974; Maitzen, 1984). These objects exhibit
strictly periodic variations in brightness, spectral profiles, and
magnetic fields; this can be explained in terms of the oblique
rotator model of a rigidly-rotating star with stable spot structures
on the surface and a stable global magnetic field that plays a
stabilizing role (Deutsch, 1970).

Anomalies in the chemical element abundances of such objects arise
as a result of long-term processes occurring in the outer calm
atmospheric layers. The observed light variations are caused by
the flux redistribution in spot structures due to the blanketing
effect in the lines and continuum. The spots of chemical anomalies
in the  light curves usually appear bright in the optical domain
and weak in the far UV, since this region contains many
bound-bound and bound-free transitions of a number of elements
(mainly Si, Fe, and rare-earth elements; see the paper by
Krti{\v{c}}ka et al. (2012) and references therein for details).
Despite the fact that mCP stars can be discovered in a fairly wide
range of parameters of the Hertzsprung--Russell (HR) diagram, the
occurrence of objects with detected magnetic fields remains almost
constant and small for all types: about 10\%, and the field
properties do not demonstrate any pronounced dependences on mass,
luminosity, or rotation (Wade et al., 2016; Sch{\"o}llerr et al.,
2017). The magnetic field is detected both in pre-MS objects
(Kholtygin et al., 2019) and those being in the final stages of
stellar evolution (Landstreet and Bagnulo, 2019). Evidence of
evolutionary variations in the magnetic field during the MS stage
 were found by Semenko et al. (2022) when studying CP stars
of the subgroups of different ages in the Orion\,OB1 association.
Unfortunately, to date, the number of the known CP stars with a
detailed modeling of the magnetic field structures is not enough
to construct a complete model of its origin and subsequent
evolution. This problem, in the authors' opinion, is extremely
interesting. To solve it, it is necessary to apply effective
criteria for selecting the mCP-star candidates for subsequent
observations with modern ground-based spectropolarimeters.

Until now, most CP stars have been identified as chemically
peculiar using spectroscopic methods or, more rarely, using the
$\Delta a$ photometry (Paunzen et al., 2005). Spectropolarimetry
was used to measure the magnetic field. As a rule, photometric
observations previously fulfilled an auxiliary role to refine the
rotation periods mainly. The things changed after publications of
the photometric archives of the Kepler, TESS, and other space
missions. High precision, sufficient completeness of the time
series and sky coverage of the surveys make it possible to use
these data for solving a wide range of astronomical problems in
addition to exoplanet detection including asteroseismological
studies and rotational modulation of mCP stars ({H{\"u}mmerich}
et~al., 2018; David-Uraz et al., 2019; {Mikulá{\v{s}}ek}).
Although the amplitude of the photometric variability of such
stars is very small and usually does not exceed $0\fm12$ in the
$V$ band, the time series are more suitable for determining the
rotation period than the spectra. The light curves of \mbox{mCP
stars} are smooth and can be fit reasonably by a single or double
wave (Mathys and Manfroid, 1985; Dukes and Adelman, 2018;
{Mikulá{\v{s}}ek} et al., 2018). They are well approximated by a
second-order polynomial which corresponds to the model of a
rotating star with one or two large photometric spots. The shape
and period of the light curves of \mbox{CP stars} persist for
decades ({{\v{Z}}i{\v{z}}{\v{n}}ovsk{\'{y}}}, 1994). In the paper by
H{\"u}mmerich et~al. (2018), based on photometry from the Kepler
satellite, several new \mbox{CP stars} were found and an
unexpected variety of shapes of their light curves was shown.
Using the spectra obtained with the 1-m Zeiss-1000 SAO RAS
telescope, observations with the 60-cm telescope of the Stara
Lesna observatory (Slovakia), and the LAMOST survey archive, the
authors of the cited paper classified the peculiarity type of the
sample objects and identified 39 new CP stars out of 46
photometric candidates (85\%). The resulting list of CP stars
became the basis for spectropolarimetric monitoring which has been
carried out with the Main Stellar Spectrograph (MSS) of SAO RAS
since 2019. In this paper, we present the first results of a
comparative analysis of the variability of a number of CP stars
that we have selected based on the photometric data from the
Kepler satellite for spectropolarimetric monitoring at the 6-m BTA
telescope and estimate their fundamental parameters.

\section{SAMPLE SELECTION AND ANALYSIS TECHNIQUE}

\subsection{Selection Criteria}
 The selection of candidates for
spectropolarimetric observations was carried out based on the
photometric data from the Kepler mission according to the
method proposed by colleagues from the Masaryk University, Brno
(H{\"u}mmerich et~al., 2018). When analyzing the time series,
first of all, it is necessary to distinguish the rotational
variability  from the variability associated with
 the $\gamma$\,Dor, $\beta$\,Cep-type pulsations, slowly
pulsating B stars, as well as with the orbital motion of the star;
therefore, when compiling the sample, we used the following
criteria:
 \begin{list}{}{\setlength\leftmargin{7mm} \setlength\topsep{1mm} \setlength\parsep{0mm}
\setlength\itemsep{1pt}}
\item [1)] the early $B$ to early $F$
spectral type with corresponding color index or effective
temperature (if available);
\item [2)] the rotation period greater
than 0$\,.\!\!^{\rm d}5$;
 \item [3)] the presence of a single frequency and
the corresponding harmonics on the periodograms;
\item [4)] the
light curve is stable or varies slightly during the entire
observation period;
 \item [5)] the variability amplitude  does not
exceed several hundredths of magnitude. \end{list}

From the resulting list, we selected 10 candidates for subsequent
spectropolarimetric monitoring with the BTA telescope. This sample
includes both new \mbox{CP star} candidates and well-known
\mbox{mCP stars} for the method checkout.

\subsection{Photometric Data} In this paper, we used the data from
the MAST\footnote{\url{https://archive.stsci.edu/}} archive
obtained by the TESS (Transiting Exoplanet Survey Satellite) space
telescope (Ricker et al., 2015) launched to search for exoplanets
with the transit method. During the two-year period of the main
program, the mission covered 85\% of the whole sky carrying out
observations in the overlapping sectors of the $96^\circ\times
24^\circ$ size. This made it possible to obtain photometric series
for more than 470 million point sources. Depending on location,
 objects were observed on different time domains: from 27.4
days to almost 1 year. TESS was launched on April 18, 2018, and in
December of the same year the data of the first two sectors became
publicly available.

We analyzed the photometric time series  using the Lafleur--Kinman
method (Lafler and Kinman, 1965). The long-term trends associated
with the technical issues caused that are inherent in the space
telescope observations were preliminarily excluded.

The rotation periods  from the TESS data are in good
agreement with those published earlier in the paper by
{H{\"u}mmerich et~al. (2018), where the photometric series of the
Kepler mission were used for searching. Small differences in the
fourth or fifth decimal place are inevitable when using different
data sets and period search methods, their study is beyond the
scope of this paper.

\subsection{Spectropolarimetry} The observations were carried out
with the MSS spectrograph (Panchuk et al., 2014) of the 6-m BTA
telescope with a circular polarization analyzer (Chountonov, 2016)
equipped with the $\lambda/4$ rotating phase plate. The \mbox{E2V
CCD42-90} \mbox{CCD} with a size of $4600\times2000$ elements was
used as a light detector. Each observation involves obtaining a
pair of the Zeeman spectra with the plate rotated by 90$^\circ$.
Such a procedure allows one to exclude instrumental polarization
and other technical issues that may cause false detection of a
magnetic field. The exposure time was chosen so that the $S/N$
ratio in the spectra was at least 100. In each observation night,
in addition to the study targets, the spectra of the standard
stars were obtained: the stars with a well-known magnetic phase
curves as well as the stars with zero magnetic fields.

The spectra were extracted using the software package written for
the {\tt ESO MIDAS} environment in SAO RAS (Kudryavtsev, 2000).
The width of the observed range was 600 \AA\  in the interval of
\mbox{4400--4900}~\AA. The choice of the range is determined by
the presence of a sufficient number of lines inside it to measure
a magnetic field with acceptable accuracy.

The magnetic field was measured according to the method proposed
by Bagnulo et al. (2002). The longitudinal field measurement error
is quite sensitive to the $S/N$ ratio, $FWHM$, and the profile
of the measured spectral lines.

The atmospheric parameters were estimated using the {\tt SME}
software for calculating synthetic spectra (Piskunov and Valenti ,
2017). By varying the effective temperature $T_{\rm eff}$, the
surface gravity $\log g$, the radial velocity $V_{R}$, the
rotation velocity projected on the line of sight $v_e \sin i$ and,
if necessary, the metallicity $[{\rm M/H}]$, we achieved the best
fit between the observed H${\beta}$ Balmer line and the synthetic
spectrum. To build the last, we used the {\tt LLmodels} model grid
(one-dimensional plane-parallel atmosphere, the LTE approximation,
the ``line-by-line'' approach to calculate the spectral line
profiles) (Shulyak et al., 2004) and a list of lines obtained from
the VALD database (Piskunov et al., 1995). Note that detailed
modeling of the chemical abundance of objects is beyond the scope
of this paper; here we restrict ourselves to only approximate
estimation of parameters from the spectra we have.

To build the phase curve of the magnetic field, we used the found
photometric periods, after which the resulting curves were
approximated by one or the sum of two sinusoids. The zero phase of
the photometric light curve corresponds to the maximum brightness
of the series.

\section{RESULTS}

We estimated the reliability of detecting the magnetic field in the studied stars using the criterion of the reduced statistics of $\chi^2/n$ calculated with the formula:

\begin{equation} \chi^2/n = \dfrac{1}{n} \sum_{i=1}^n \left(\dfrac{B_{i}}{\sigma_{i}}\right)^{2}, \nonumber\end{equation}
where $B_i$ and $\sigma_i$~are individual measurements of a magnetic field and the corresponding errors. Traditionally, as for our method (see, e.g., Romanyuk et al. (2019)), we will assume that the magnetic field is reliably detected at \mbox{$\chi^2/n > 5$}.

The results of individual magnetic measurements of the sample objects are given in Table~\ref{tab:log1}, where JD is a Julian date of observations, $B_e \pm \sigma$ is the estimation of the longitudinal magnetic field and the corresponding root-mean-square error.
Below we give the comments on the results obtained during the study.

\subsection{KIC\,4180396 = HD\,225728} Analysis of the TESS
photometric data has shown that the observations of the object are
best described by the following ephemeris: $$ {\rm JD} =
2455684.20056 + 3.68435 E. $$ Figure~1a shows the resulting light
curve. It exhibits a double wave with the primary maximum and
minimum at the rotation period phases: \mbox{$\phi = 0.0$} and
\mbox{$\phi = 0.28$}. They are followed by the secondary maximum
and minimum at the phases: $\phi = 0.52$ and \mbox{$\phi = 0.78$}
respectively. The full amplitude of the light variation is
equal to \mbox{$\Delta m = 0\fm013$}.
The positive magnetic extremum  \mbox{$B_e{\rm (max)} = 1520\,\pm\,52$}~G, corresponds to the photometric maximum.
The negative extremum of the field is slightly shifted relative to the secondary photometry maximum towards the primary minimum and is in the phase $\phi = 0.2$. The field is best approximated by a double sine curve (Fig.~1b), while the secondary harmonics are almost invisible. The estimate \mbox{$\chi^2/n = 319.8$} confirms reliable detection of the magnetic field.

As a result~ of the~ spectrum~ approximation (see Fig.~1c) the following atmospheric parameters are defined: $T_{\rm eff}=11\,283$~K, $\log g=3.80$, \linebreak \mbox{$v_e \sin i = 28.67$}\,km\,s$^{-1}$, \mbox{$V_{R}=-8.65$~km s$^{-1}$,} and \linebreak
\mbox{$[{\rm M/H}]=-0.096$}.
\subsection{KIC\,5264818 = HD\,180374}

The best ephemeris that we obtained as a result of the photometric series analysis:
 $$ {\rm JD} = 2458683.9789 + 1.90291 E. $$

\LTcapwidth=\columnwidth
\renewcommand{\baselinestretch}{0.75}
\setlength{\tabcolsep}{5.5pt}
 \begin{center}
    \begin{longtable}{c|c| r@{$\,\pm\,$}l}
    \caption{Longitudinal magnetic fields of the sample stars according to our measurements}\label{tab:log1}\\
    \hline
    Star & JD 2450000+ & \multicolumn{2}{c}{$B_e \pm \sigma$, G} \\
    \hline
    \endfirsthead
    \caption{(Continued) }\\
    \hline
    Star & JD 2450000+ & \multicolumn{2}{c}{$B_e \pm \sigma$, G} \\
    \hline
    \endhead
 \hline
    \endfoot
    \hline
    \endlastfoot
KIC\,4180396 & 8578.553  & $-$544 & 70  \\
& 8603.478  & $-$650 & 75  \\
& 8777.274  & $-$578 & 70  \\
& 8802.198  & $-$250 & 80  \\
& 8805.222  &   1517 & 80  \\
& 9033.494  &   1400 & 70  \\
& 9099.448  &   1520 & 70  \\
\hline
KIC\,5264818 & 8597.496  &   1004 & 65  \\
& 8621.370  & $-$995 & 70  \\
& 8624.414  &    227 & 70  \\
& 8777.211  & $-$810 & 80  \\
& 8778.165  &    940 & 70  \\
& 8921.519  &    300 & 50  \\
& 9001.340  &    960 & 50  \\
& 9006.387  &    101 & 65  \\
& 9032.509  & $-$566 & 50  \\
& 9061.482  & $-$629 & 50  \\
& 9097.388  & $-$757 & 55  \\
\hline
KIC\,5473826 & 8600.401  & $-$701 & 203 \\
 & 8801.276  &     99 & 205 \\
 & 8805.194  & $-$286 & 120 \\
 & 9031.340  & $-$479 & 180 \\
 & 9096.346  & $-$147 & 143 \\
 \hline
KIC\,6065699 & 8577.541  &    650 & 80  \\
 & 8620.381  &    563 & 50  \\
 & 8621.438  &    550 & 122 \\
 & 8624.481  &    463 & 80  \\
 & 8777.330  &    731 & 60  \\
 & 8778.336  &    676 & 70  \\
 & 8799.281  &    781 & 63  \\
 & 8802.254  &    720 & 46  \\
 & 8805.149  &    820 & 60  \\
 & 9006.424  &    539 & 70  \\
 & 9060.308  &    445 & 70  \\
 & 9061.535  &    691 & 74  \\
 & 9102.403  &    845 & 90  \\
 \hline
KIC\,6278403 & 8601.473  & $-$220 & 90  \\
 & 8603.420  & $-$179 & 90  \\
 & 8620.434  & $-$130 & 90  \\
 & 8621.411  &    171 & 90  \\
 & 8624.453  &     63 & 90  \\
 & 9006.365  &     10 & 90  \\
 & 9097.367  &  $-$20 & 90  \\
 & 9431.500  &    100 & 100 \\
 \hline
KIC\,6864569 & 8758.312  & $-$195 & 70  \\
 & 8778.297  &     16 & 80  \\
 & 8830.141  & $-$305 & 85  \\
 & 9060.502  & $-$403 & 53  \\
 & 9455.439  &    177 & 91  \\
 \hline
KIC\,8161798 & 8600.490  &  $-$50 & 100 \\
 & 8802.151  &  $-$55 & 100 \\
 & 8830.205  &    250 & 100 \\
 & 9032.464  &    240 & 100 \\
 & 9060.416  & $-$157 & 100 \\
 & 9096.433  &    118 & 100 \\
 & 9336.542  &    210 & 100 \\
 & 9455.347  & $-$105 & 100 \\
 \hline
KIC\,8324268  & 8577.516  &    135 &  90 \\
 & 8600.558  &   $-$7 &  90 \\
 & 8601.559  &  $-$11 &  90 \\
 & 8620.353  & $-$450 &  90 \\
 & 8624.504  & $-$185 &  90 \\
 & 8777.352  &      0 &  90 \\
 & 8778.360  &    460 &  90 \\
 & 8799.233  & $-$190 &  90 \\
 & 8801.326  & $-$170 &  90 \\
 & 8805.305  & $-$227 &  90 \\
 & 9000.368  & $-$117 &  90 \\
 & 9021.410  &  $-$20 &  90 \\
 & 9096.534  & $-$150 &  90 \\
 \hline
KIC\,10324412  & 8620.323  &     50 & 55  \\
 & 8621.336  &      2 & 70  \\
 & 8624.377  & $-$101 & 70  \\
 & 8799.259  & $-$103 & 90  \\
 & 8801.153  &   $-$5 & 60  \\
 & 8976.335  &  $-$65 & 70  \\
 & 9061.456  &  $-$23 & 170 \\
 & 9097.416  & $-$164 & 122 \\
 & 9099.397  &  $-$72 & 80  \\
 & 9455.276  & $-$138 & 127 \\
 \hline
KIC\,11560273  & 8758.257  &      0 & 100 \\
 & 8777.375  &  $-$48 &  90 \\
 & 8778.387  &  $-$17 & 100 \\
 & 8830.260  &     40 & 100 \\
 & 9061.532  &     56 & 100 \\
    \end{longtable}
\end{center}
\renewcommand{\baselinestretch}{1.0}
\begin{figure} 
\includegraphics[width=0.99\linewidth, bb = 0 0 445 220,clip]{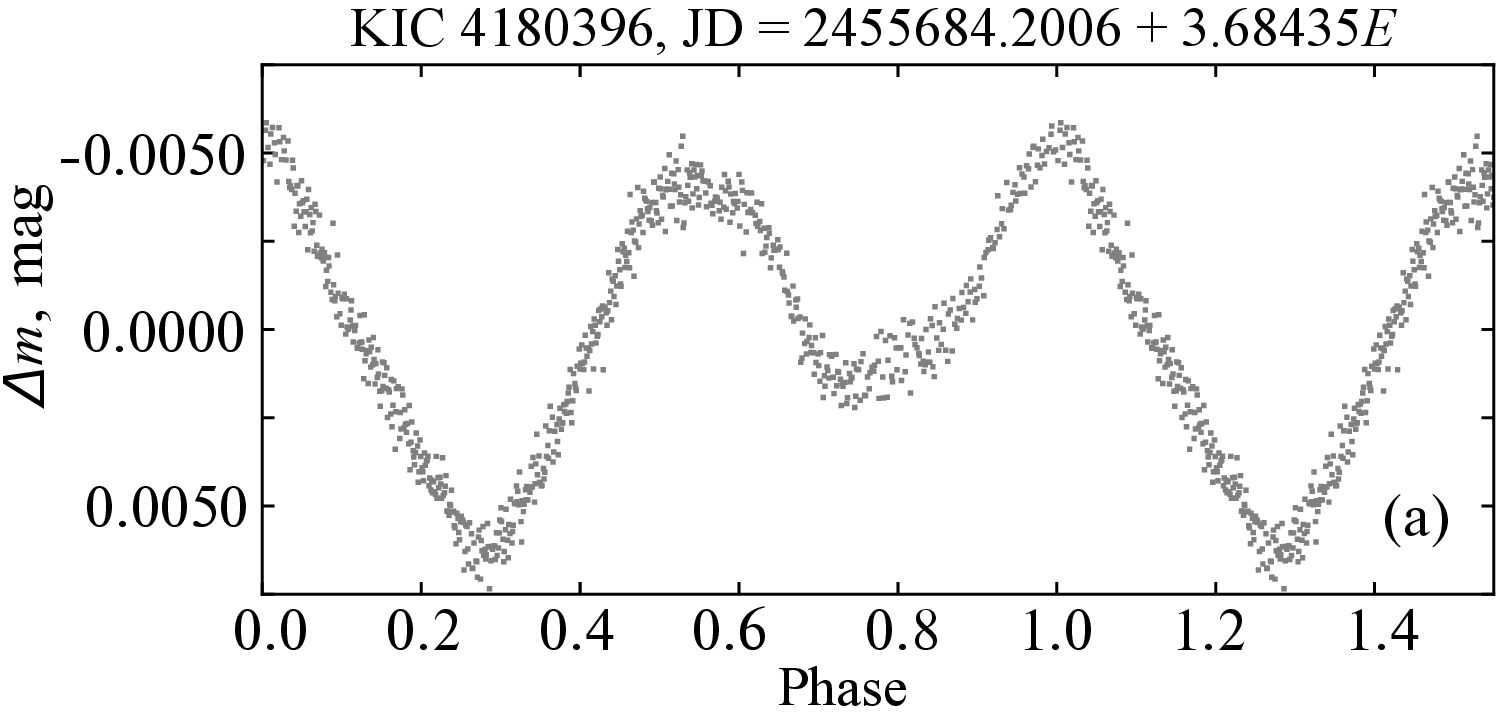} \includegraphics[width=0.96\linewidth, bb = -10 0 580 370,clip]{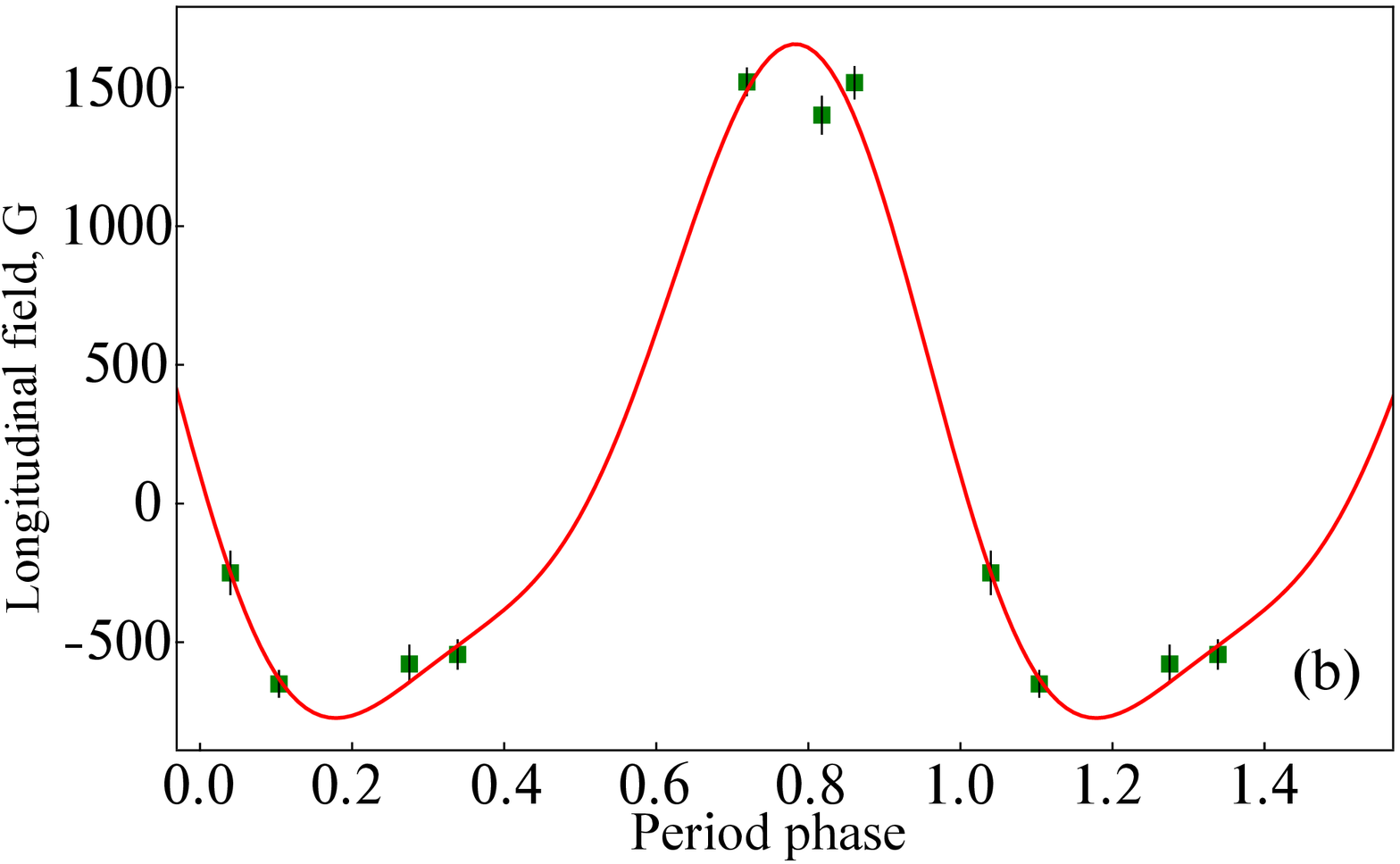}\\ \includegraphics[width=0.97\linewidth, bb = 0 0 620 405,clip]{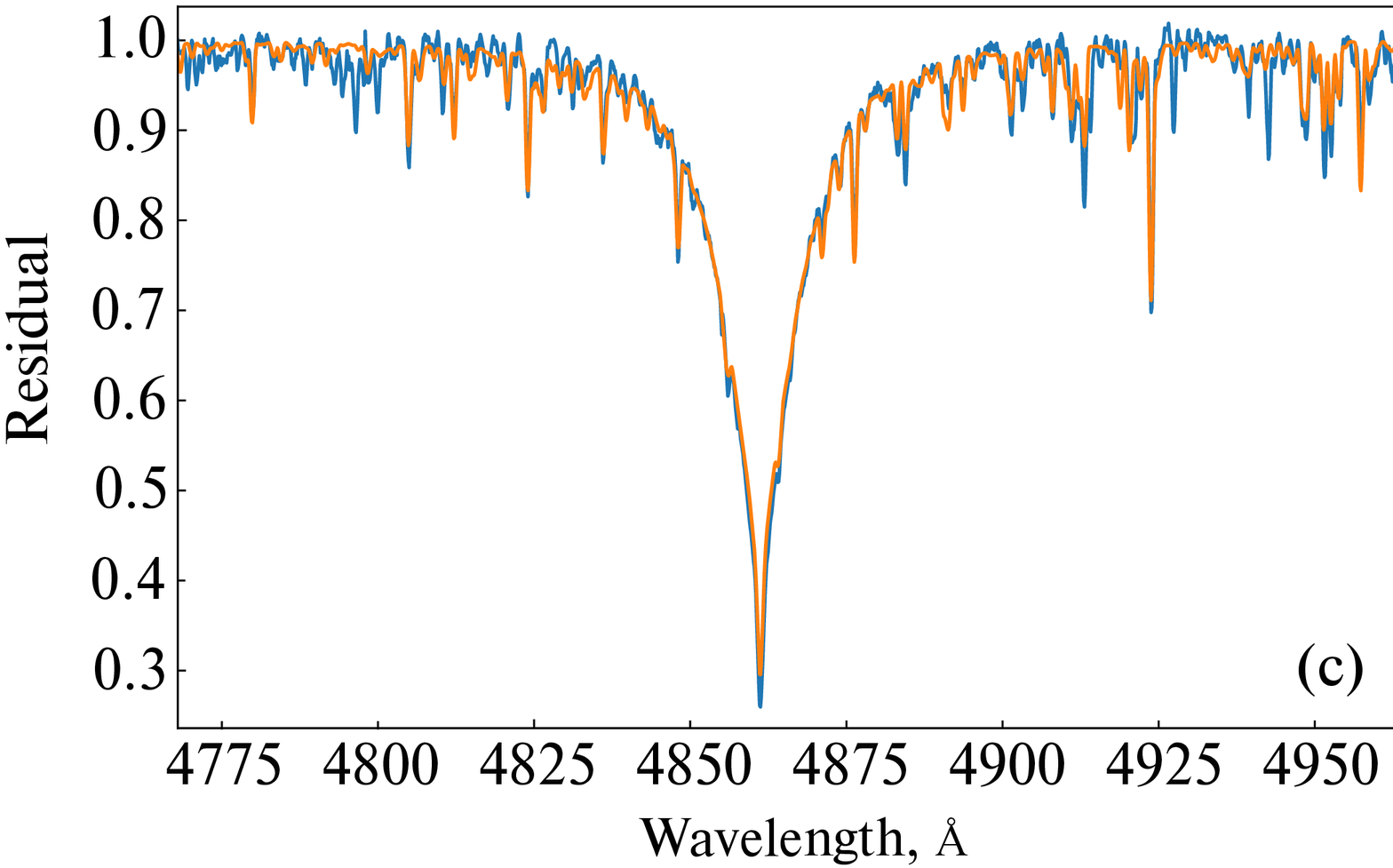} \caption{Analysis of the star KIC\,4180396. Panel (a) shows the light curve according to the TESS satellite data; panel (b) shows the phase curve of the magnetic field; panel (c) presents the modeling result of the H${\beta}$ line, observations are shown in blue, and modeling results in orange.} \label{fig1} \end{figure}

\begin{figure}[t] 
 \includegraphics[width=0.99\linewidth, bb = 0 0 445 220,clip]{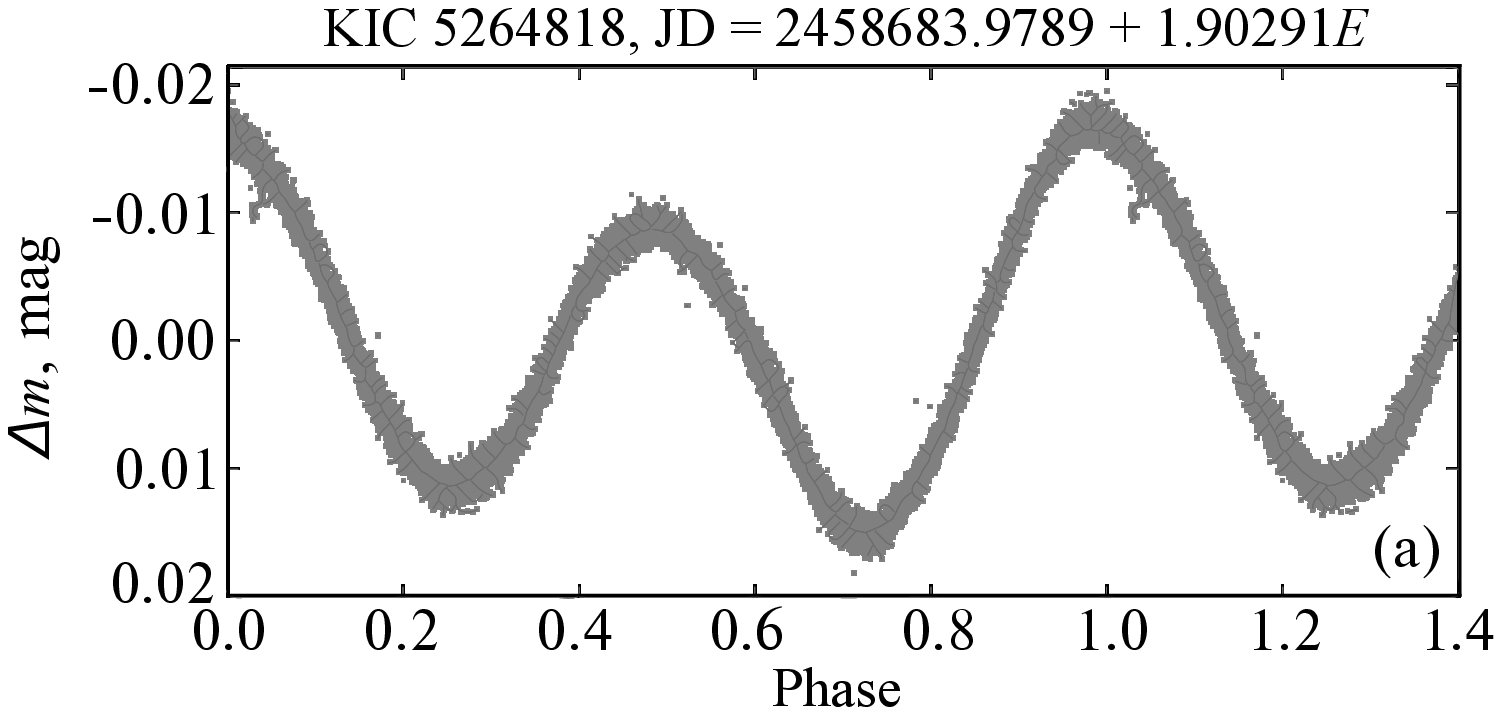} \\ \includegraphics[width=0.96\linewidth, bb = 0 0 590 375,clip ]{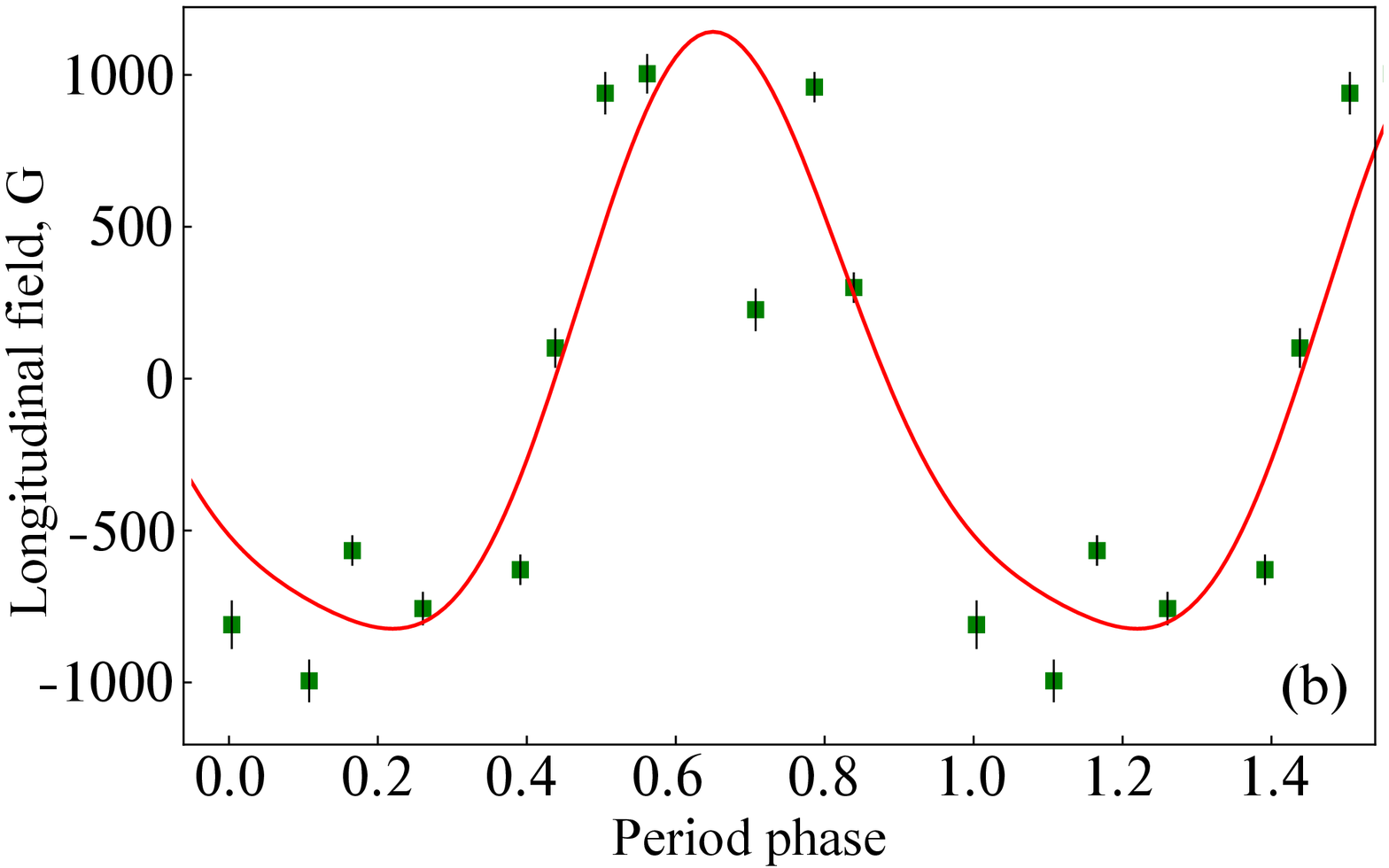}\\ \includegraphics[width=0.94\linewidth, bb = 0 0 620 410,clip]{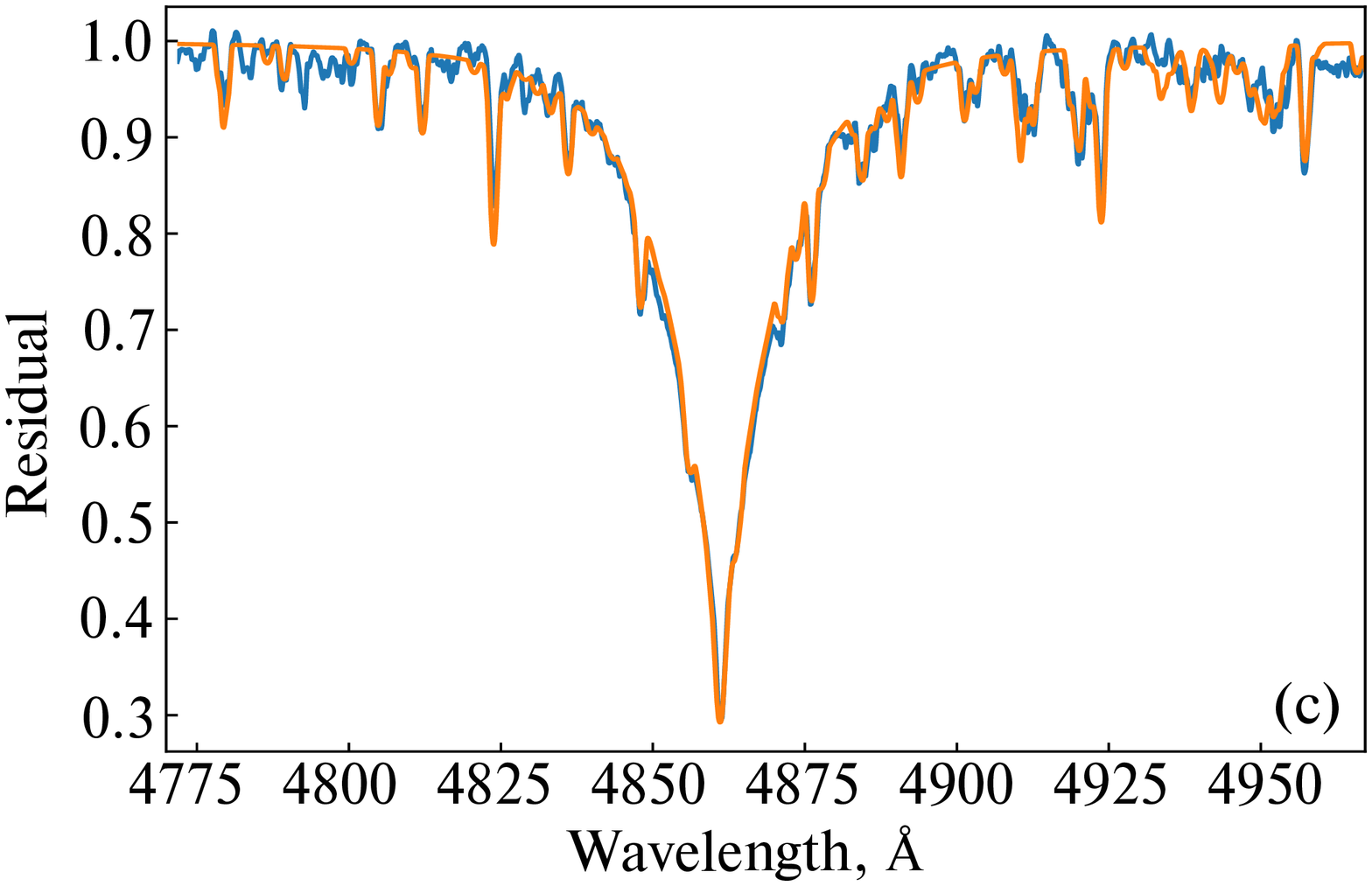} \caption{Same as in Fig.~1, for KIC\,5264818. } \label{fig2} \end{figure}

The light curve (Fig.~2a) has a more pronounced double wave. The
primary maximum is followed by the secondary minimum and maximum
in the phases: $\phi = 0.25$ and $\phi = 0.50$. Further, the
primary minimum is in the region of the phase $\phi = 0.73$. The
full amplitude is equal to $\Delta m = 0\fm037$.

The positive magnetic field extremum lies in the region of the primary minimum of the light curve; it is slightly shifted relative to it in the phase \mbox{$\phi = 0.61$}. The negative field maximum is flat and located between the primary maximum and the secondary minimum of photometry in the range of the phases \mbox{$\phi:~0.0$--$0.22$}. The magnetic field is best approximated by a double sine curve
(see Fig.~2b). The star KIC\,5264818 is reliably magnetic (\mbox{$\chi^2/n = 146.9$}).

With the approximation of the spectrum~ shown in Fig.~2c, the following~ atmospheric~ parameters~ are~ obtained:
 $T_{\rm eff}=9348$~K, $\log g=3.59$, \linebreak \mbox{$v_e \sin i = 57.37$~km s$^{-1}$}, \mbox{$V_R=-14.29$~km s$^{-1}$,} and \linebreak \mbox{$[{\rm M/H}]=-0.000$}.

 \subsection{KIC\,5473826 = HD\,226339} \begin{figure} \includegraphics[width=0.99\linewidth, bb = 0 0 445 220,clip]{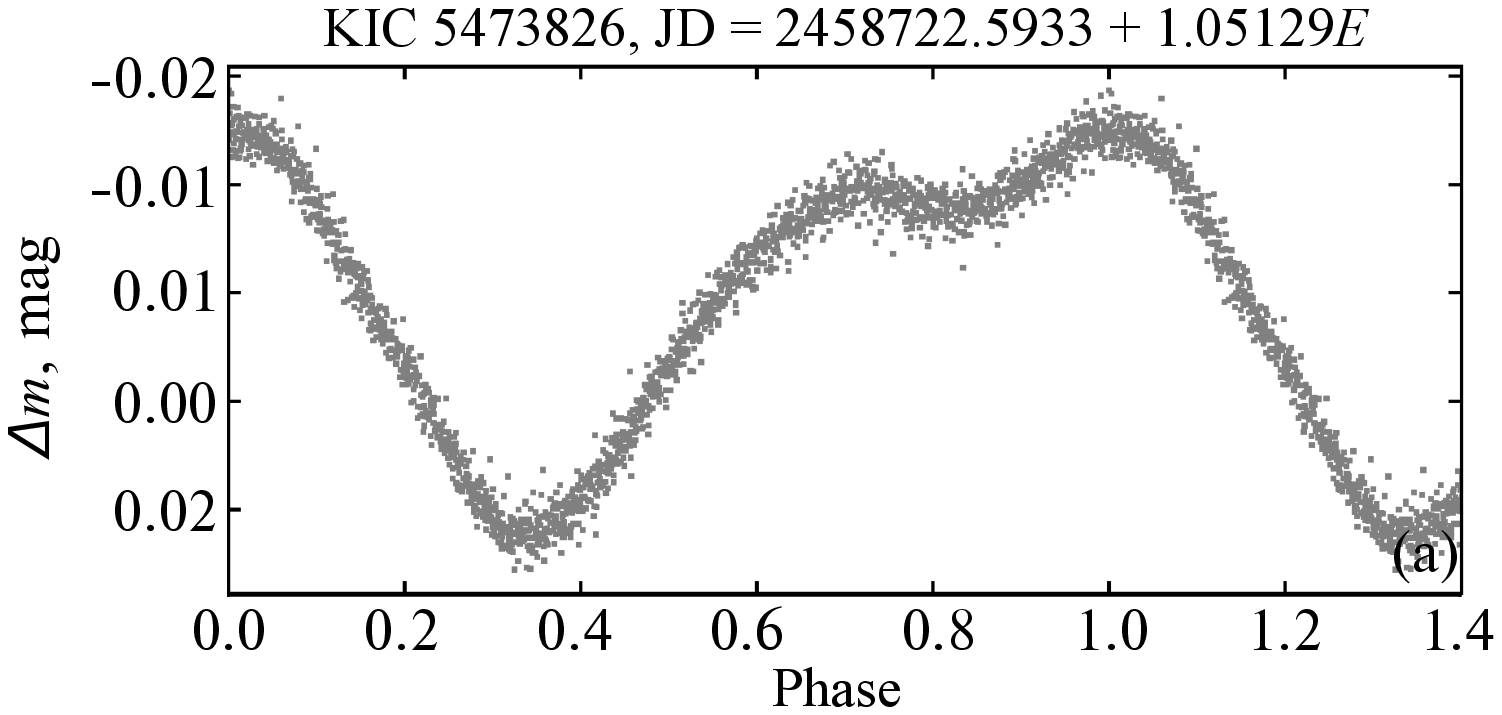} \\ \includegraphics[width=0.96\linewidth, bb = 0 0 590 375,clip ]{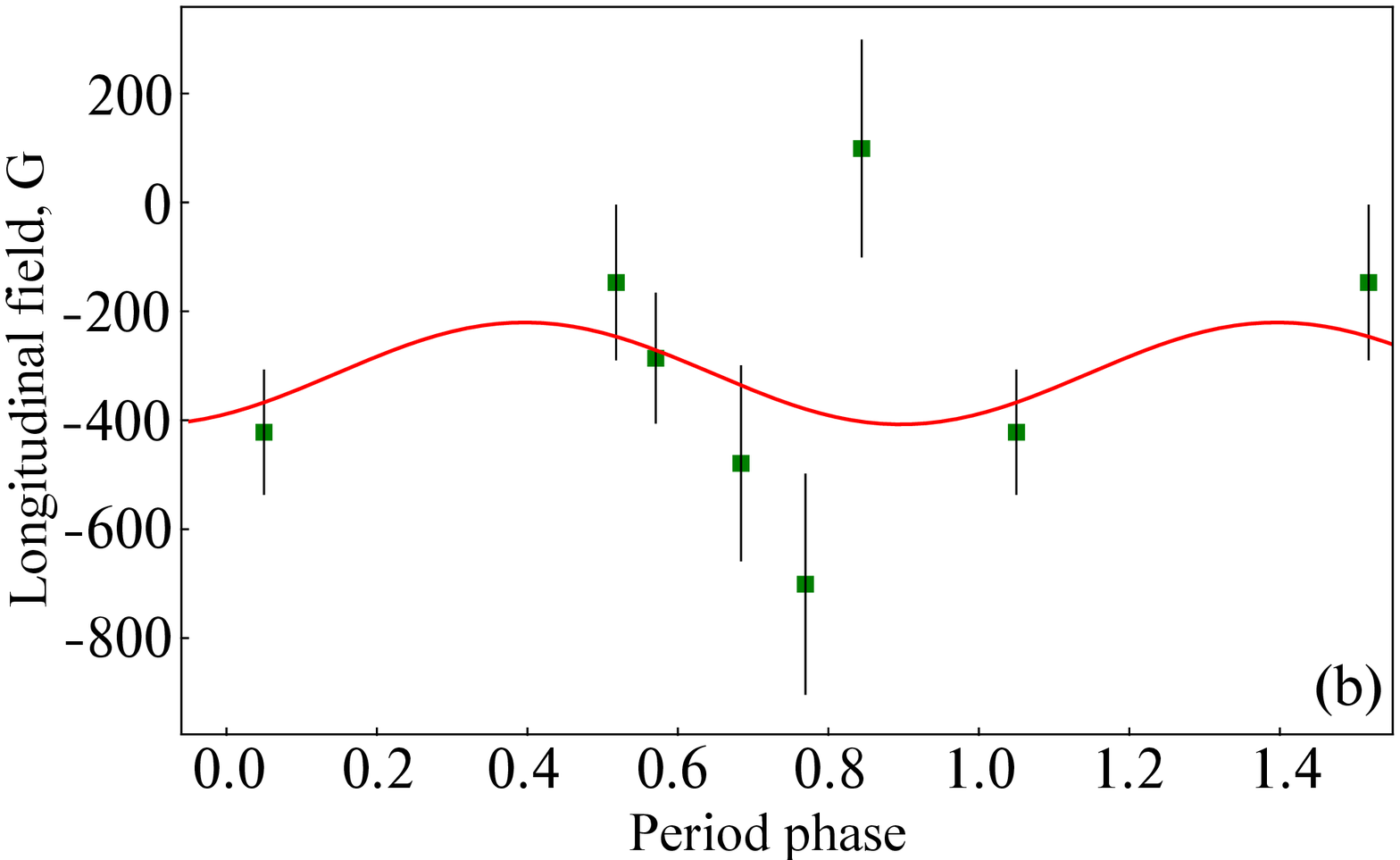}\\ \includegraphics[width=0.94\linewidth, bb = 0 0 620 410,clip]{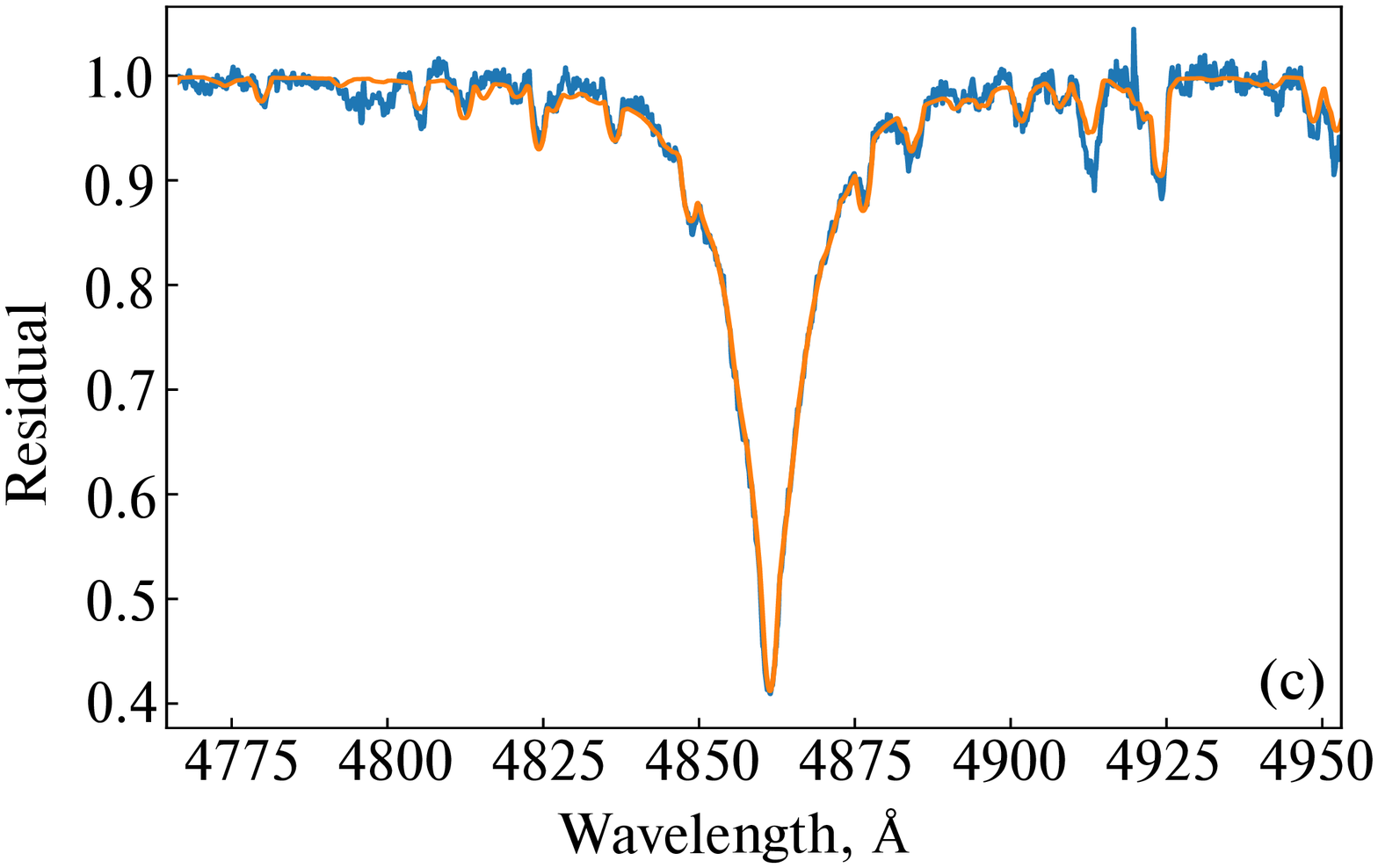} \caption{Same as in Fig.~1, for KIC\,5473826.} \label{fig3} \end{figure}

The best matched ephemeris is $$ {\rm JD} = 2458722.5933 + 1.05129
E. $$ The resulting light curve (Fig.~3a) is characterized by
weakly pronounced secondary extrema comparable to the primary
maximum. The primary minimum is in the phase $\phi = 0.34$, the
secondary maximum and minimum are in the phases: $\phi = 0.70$ and
\mbox{$\phi = 0.80$} respectively. The  full amplitude of the brightness
variation $\Delta m = 0\fm042$.

Despite the fact that the statistical criterion indicates reliable detection ($\chi^2/n = 6.62$), the magnetic field is determined with a large error due to wide and non-Gaussian line profiles in the spectrum. The only field measurement with a positive sign is in the region of the secondary minimum of the light curve. Moreover, very close, in the phase of the secondary maximum, there is a negative extremum of the field. The observed data for more reliable approximation of the magnetic curve are not enough. The study of this target will be continued.

When approximating the spectrum (Fig.~\ref{fig3}c), the following atmospheric parameters were obtained: \mbox{$T_{\rm eff}=12\,479$~K,} $\log g=3.82$, \mbox{$v_e \sin i = 83.39$~km s$^{-1}$}, and \mbox{$V_{R}=5.69$~km s$^{-1}$}.

\subsection{KIC\,6065699 = HD\,188101}

\begin{figure} \includegraphics[width=0.99\linewidth, bb = 0 0 445 220,clip]{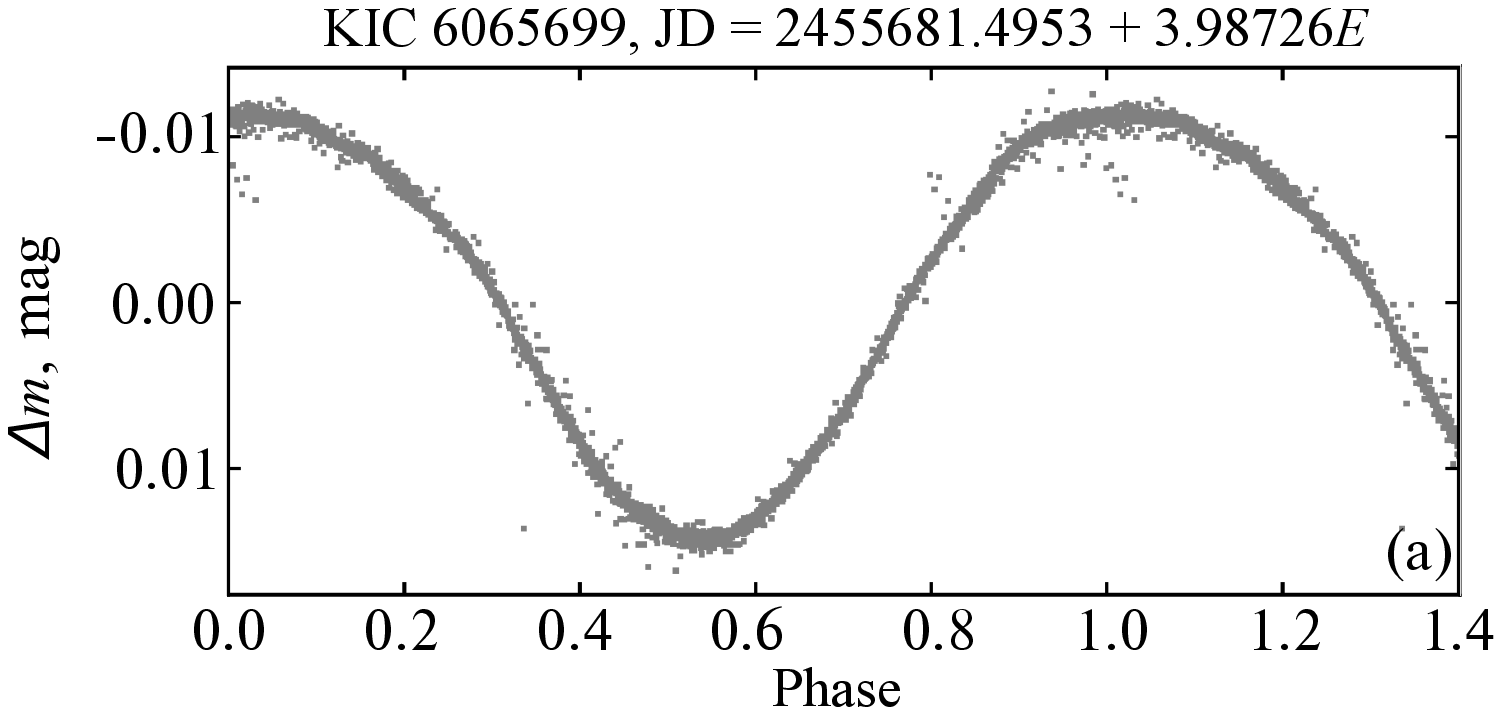} \\ \includegraphics[width=0.96\linewidth, bb = 0 0 590 375,clip ]{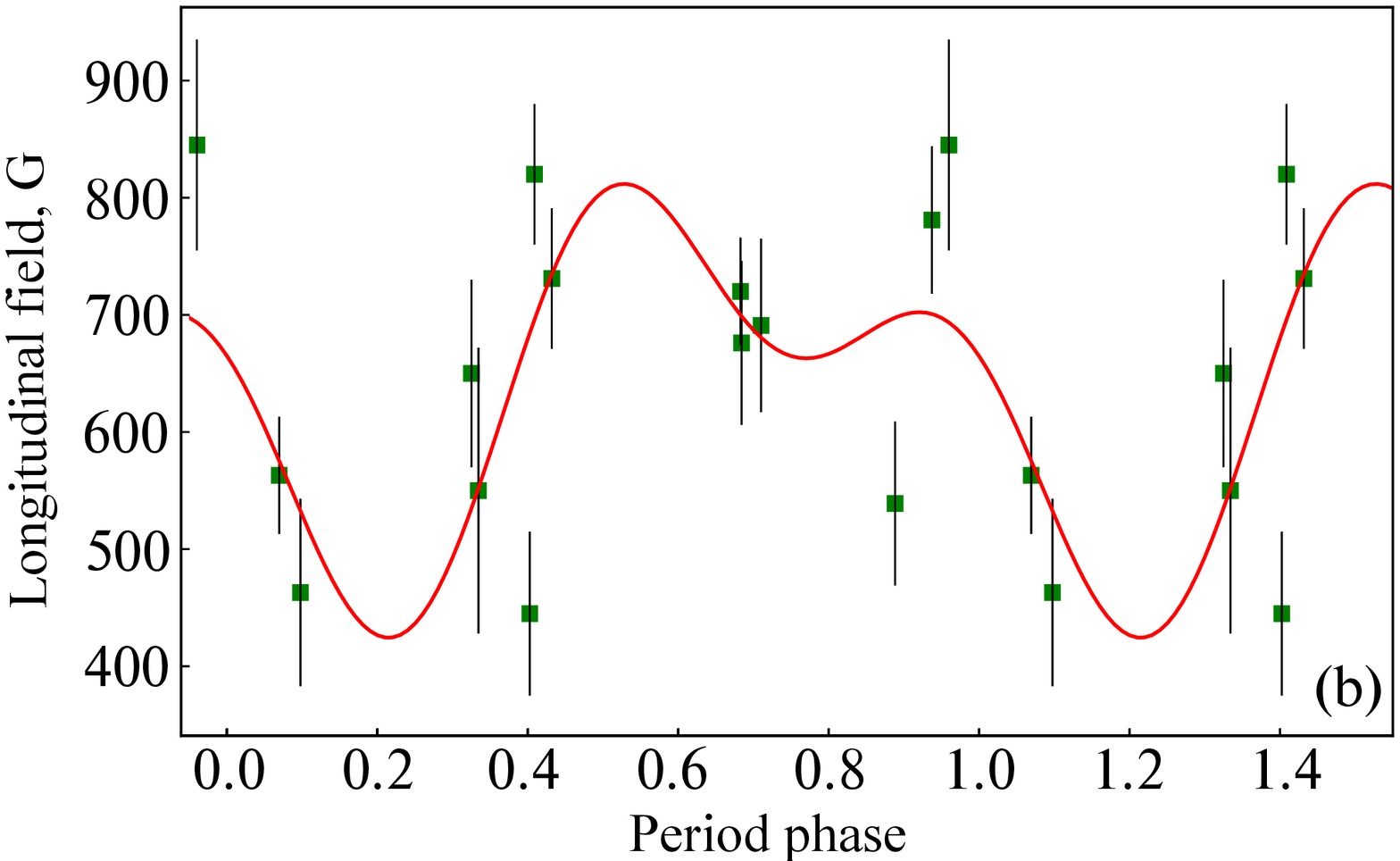}\\ \includegraphics[width=0.94\linewidth, bb = 0 0 620 410,clip]{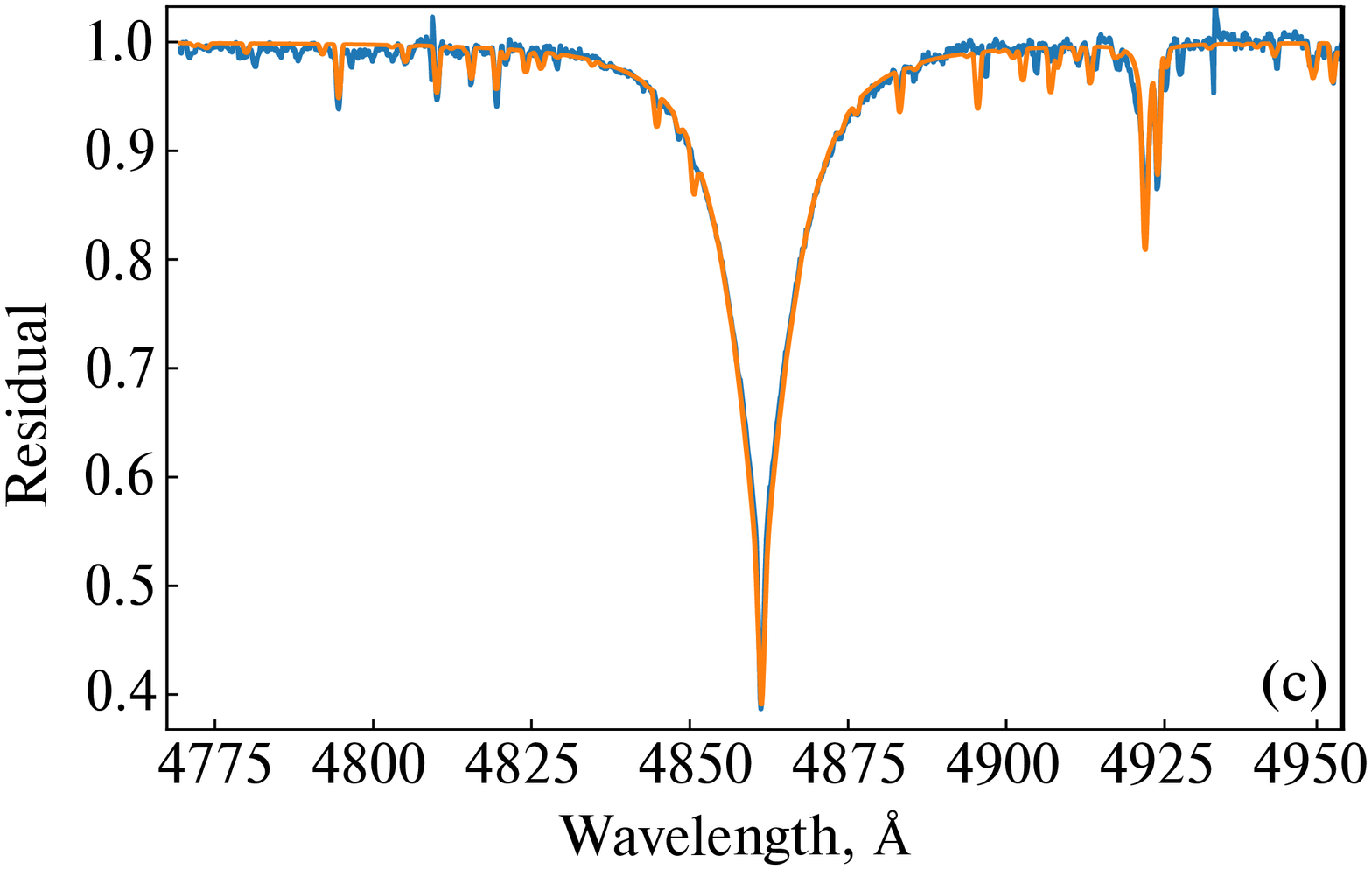} \caption{Same as in Fig.~1, for KIC\,6065699.} \label{fig4} \end{figure}

The best ephemeris:$$ {\rm JD} = 2455681.4953 + 3.98726 E.$$

The light curve is a smooth almost harmonic sine curve with the maximum in the phase $\phi = 0.0$ and the minimum in the phase $\phi = 0.55$ (Fig.~4a). The total amplitude of the brightness variation $\Delta m = 0\fm022$.

The magnetic fields, on the contrary, show large scatter and are difficult to approximate, despite the large number of measurements that have only positive polarity (see Fig.~4b). These features can be explained by the fact that the positive pole of the dipole is slightly deviated from the rotation axis of the star and directed towards the observer. This explanation is supported by the narrow spectral line profiles which are instrumental actually. According to \textbf{ $\chi^2/n = 103.75$}, the star KIC\,6065699 is magnetic.

Based on the spectrum approximation shown in Fig.~4c, the following parameters are obtained: \mbox{$T_{\rm eff}\!=\!14\,771$}~K, \mbox{$\log g\!=\!3.83$}, $v_e \sin i = 33.40$~km s$^{-1}$, $V_{R}=-1.93$~km s$^{-1}$, and $[{\rm M/H}]=-0.006$ (see Fig.~4).

When approximating the observed spectrum with a synthetic one, two problems arose.

The first is related to the impossibility of approximating the Si\,II and Si\,III lines by a model with one temperature. Good approximation of Si\,II required a model with the temperature $T_{\rm eff} = 15\,200$~K, and for Si\,III---\mbox{$T_{\rm eff}=17\,200$~K}. Most likely, this difference is caused by the stratification of this element in the atmosphere of the star.

The second problem is the impossibility of approximating the forbidden
line He\,I $\lambda$~4471. In this case, difficulties arise due to the
fact that this line can not be described using the LTE approach,
which we used in our work. For correct approximation of this line,
it is necessary to calculate a non-LTE model.

\subsection{KIC\,6278403 = HD\,181436} The best ephemeris according to the photometric data: $${\rm JD} = 2458694.6385 + 1.19114 E.$$

The light curve is flat, as can be seen in Fig.~5a, the brightness
fading lasts longer than brightening. The maximum phase $\phi =
0.0$, the minimum---\mbox{$\phi = 0.6$}. Some bend is observed at
the fading stage in the region of $\phi = 0.25$. The amplitude of
the curve \mbox{$\Delta m = 0\fm008$}.

Due to large measurement errors, we can not confirm the reliable
detection of a field ($\chi^2/n = 1.21$). According to our data,
it varies weakly and does not exceed several hundred gauss in the
absolute magnitude. Measurement errors also do not allow us to
confidently carry out the approximation. The maximum of the
obtained magnetic curve shown in Fig.~5b is positive and shifted
from the photometric minimum.

The best fit between the observed and synthetic spectra is
achieved with the parameters: \linebreak \mbox{$T_{\rm eff}=11\,261$~K}, \mbox{$\log
g=4.12$}, \mbox{$v_e \sin i = 46.5$~km s$^{-1}$}, and
\mbox{$V_{R}=-12.1$~km\,s$^{-1}$}.

\begin{figure} \includegraphics[width=0.99\linewidth, bb = 0 0 445 220,clip]{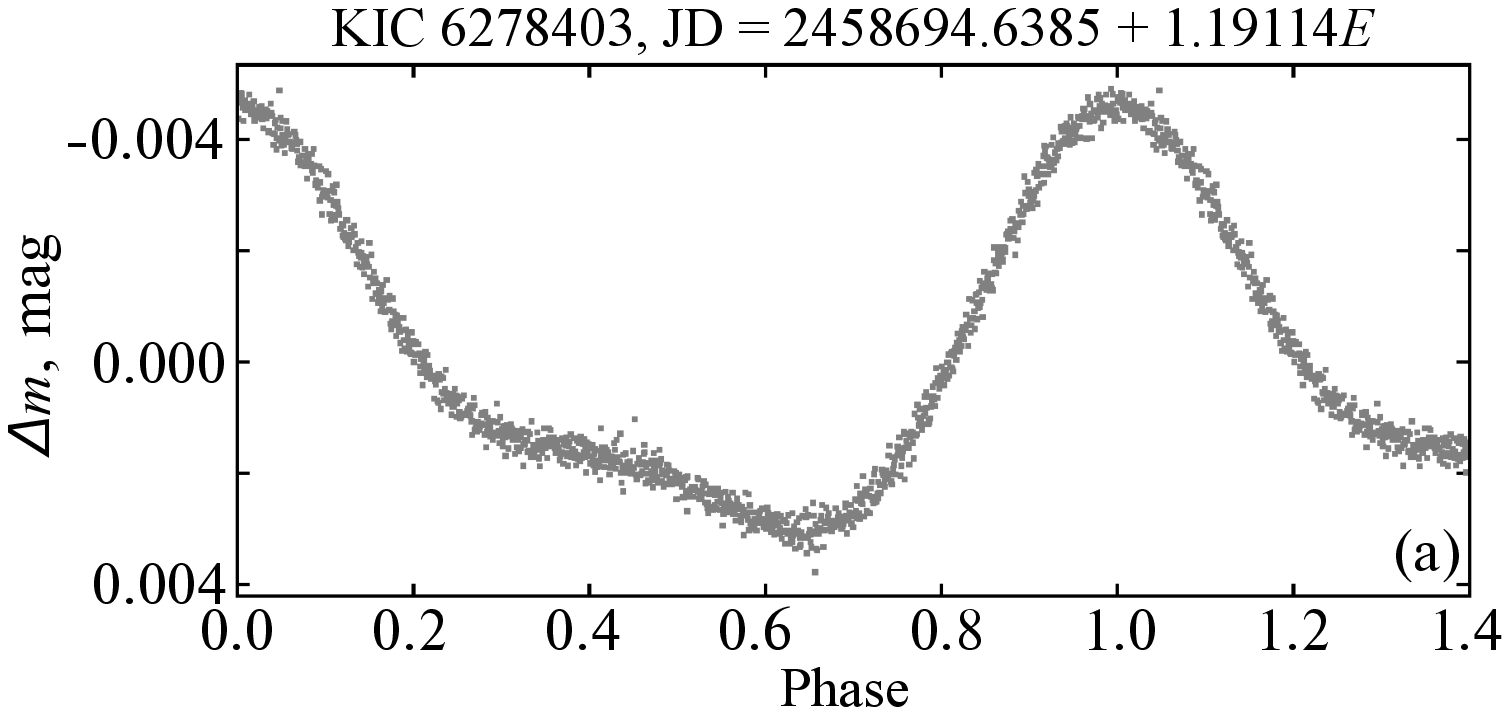} \\ \includegraphics[width=0.96\linewidth, bb = 0 0 590 375,clip ]{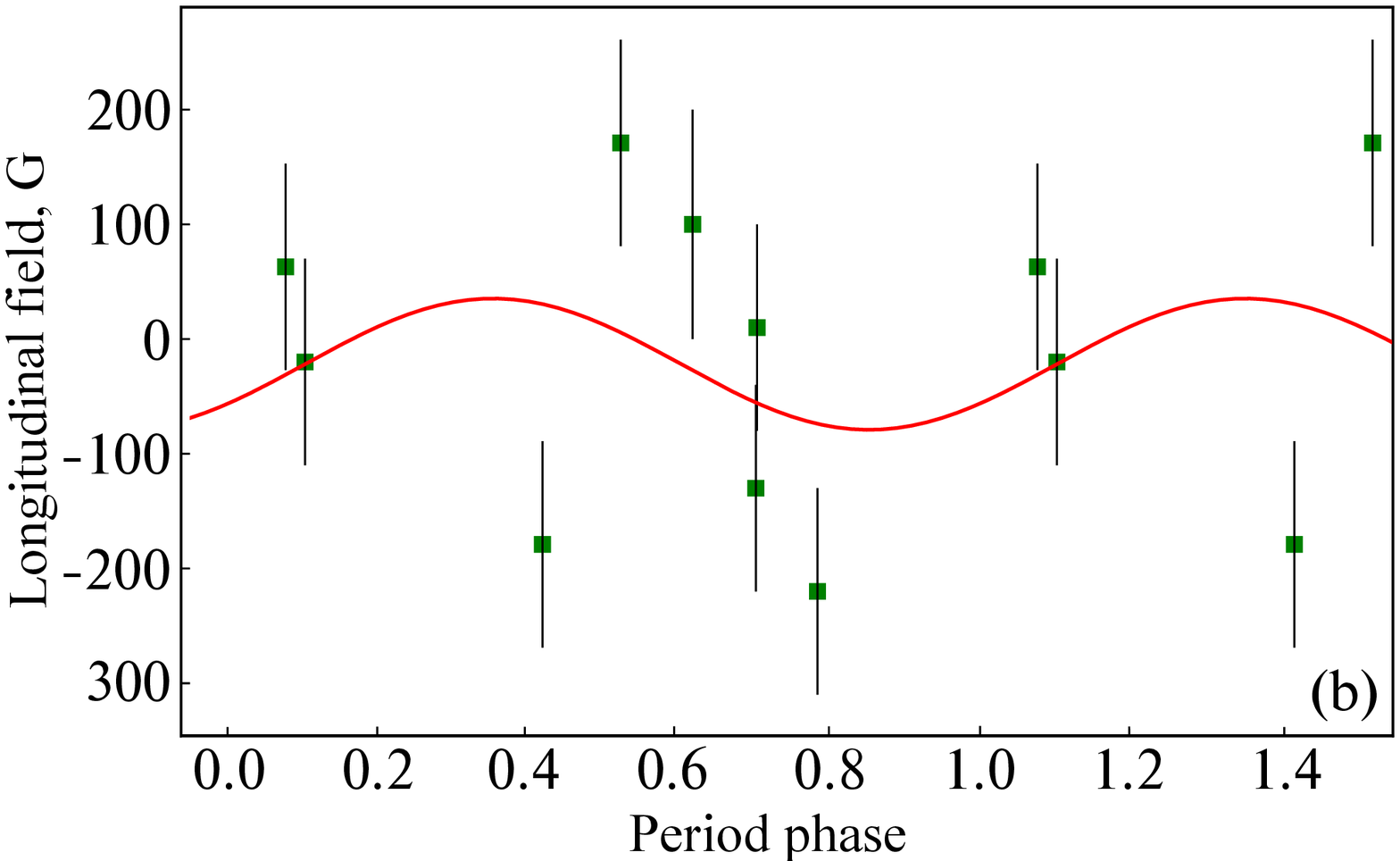}\\ \includegraphics[width=0.94\linewidth, bb = 0 0 620 410,clip]{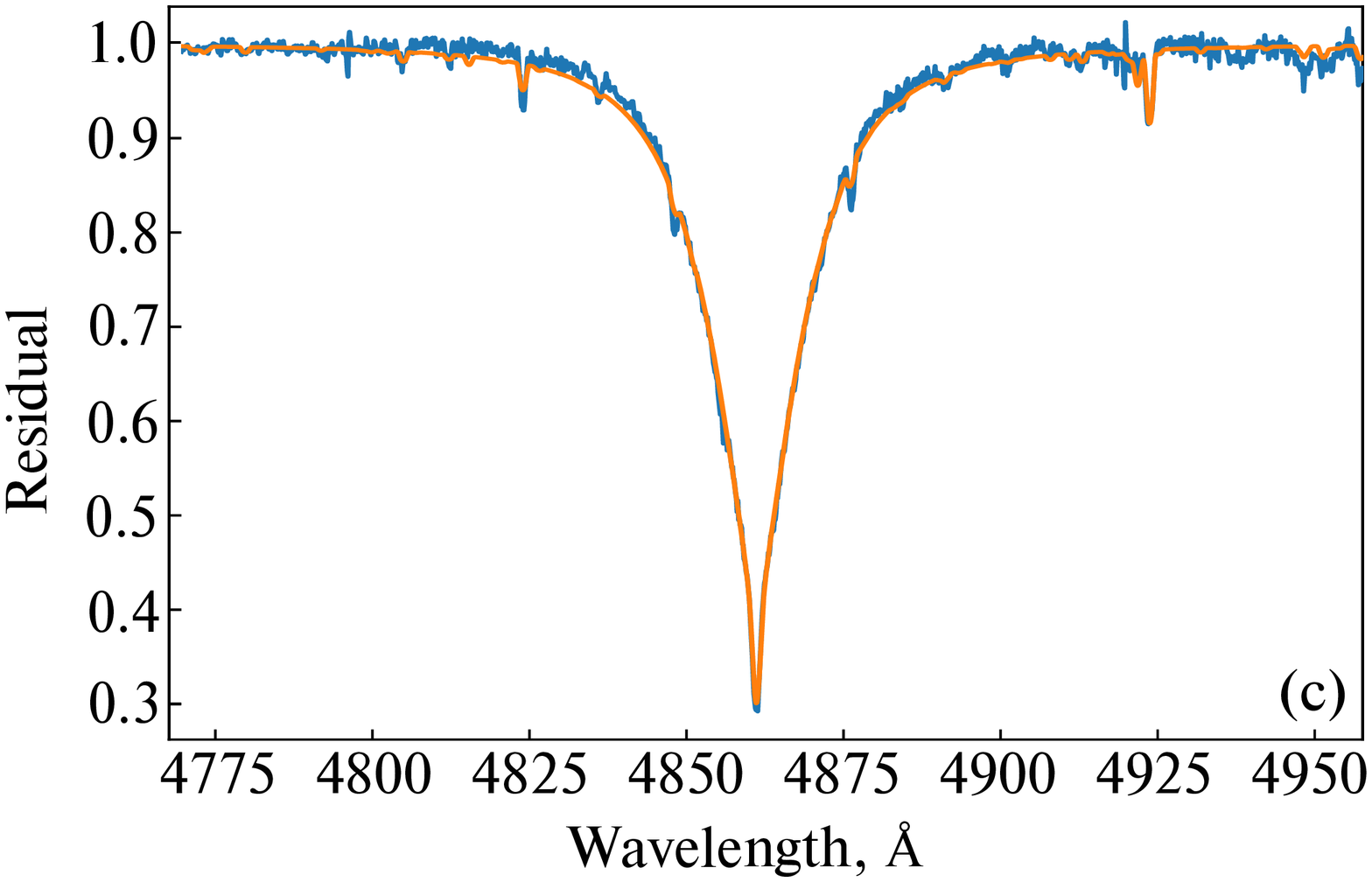} \caption{Same as in Fig.~1, for KIC\,6278403.} \label{fig5} \end{figure}

\subsection{KIC\,6864569 = BD+42$^{\circ}$3356}

\begin{figure}[t] \includegraphics[width=0.99\linewidth, bb = 0 0 445 220,clip]{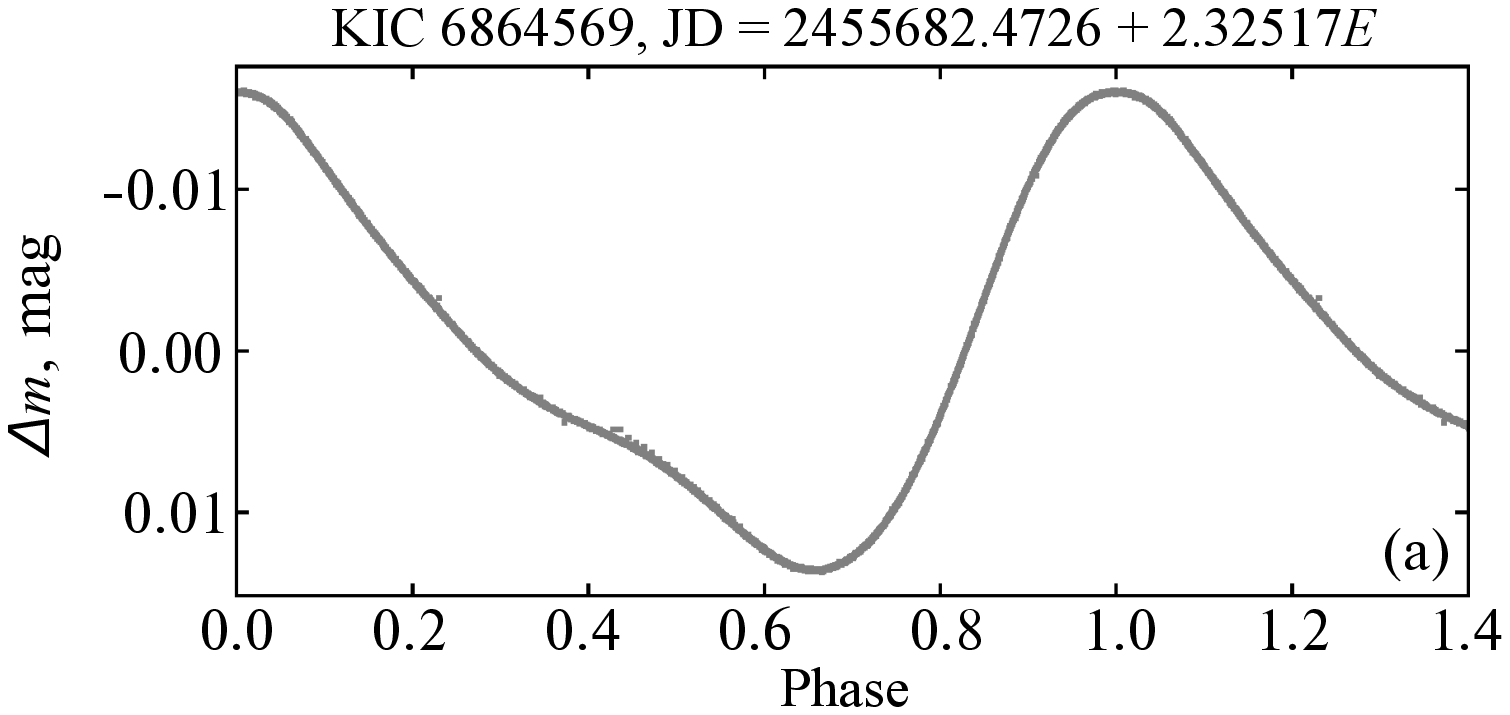} \\ \includegraphics[width=0.96\linewidth, bb = 0 0 590 375,clip ]{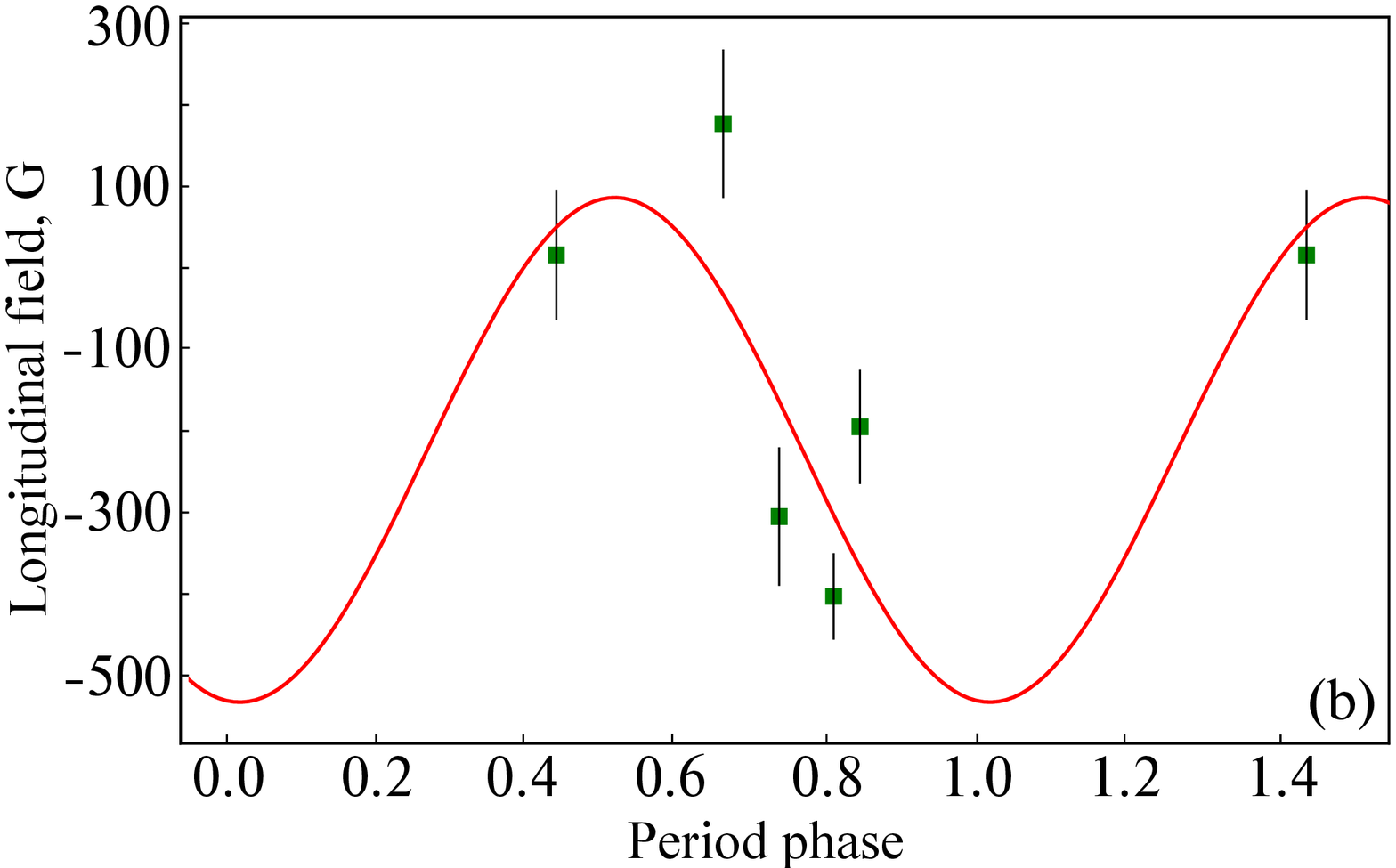}\\ \includegraphics[width=0.94\linewidth, bb = 0 0 620 410,clip]{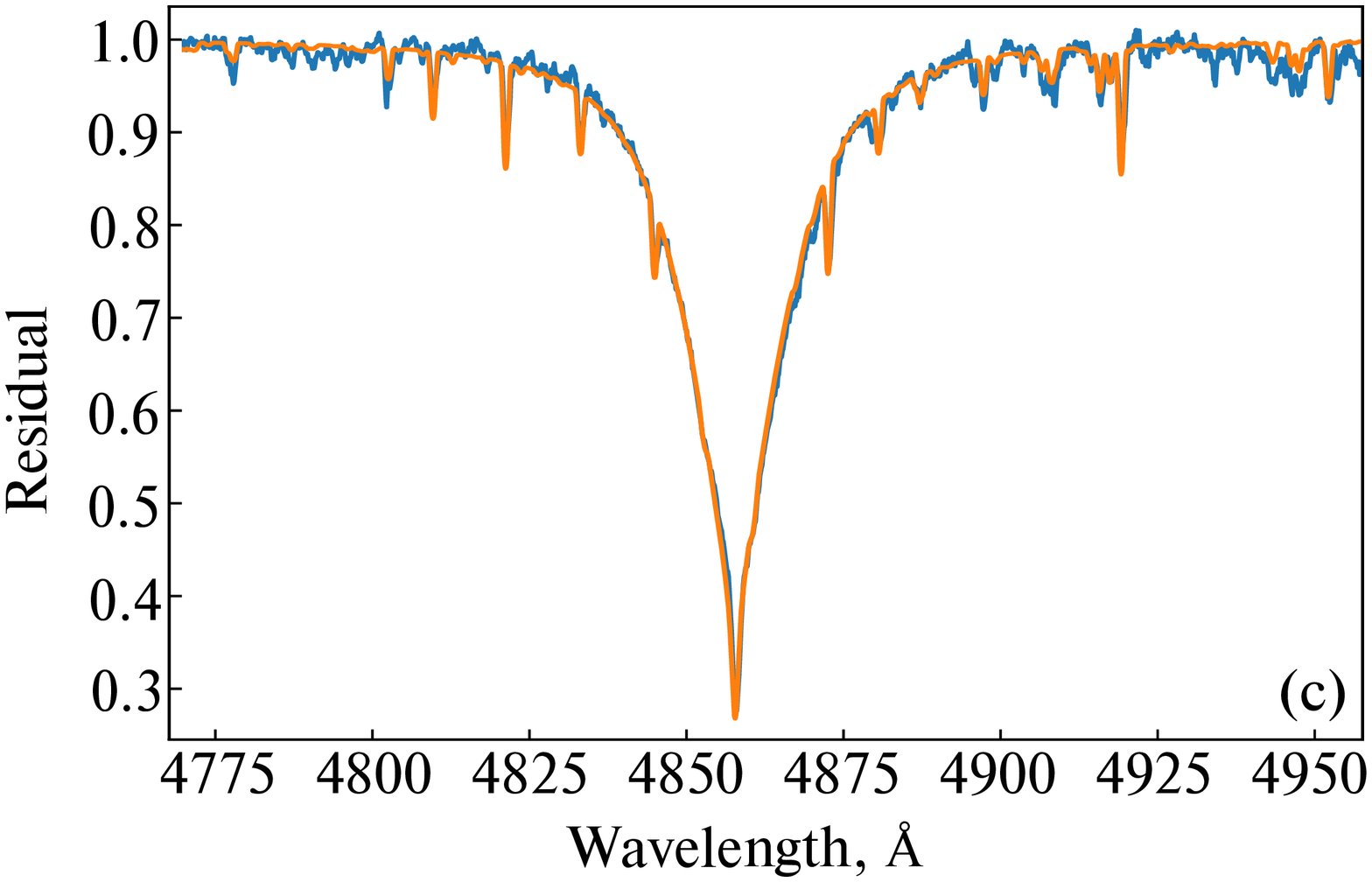} \caption{Same as in Fig.~1, for KIC\,6864569.} \label{fig6} \end{figure}

The best ephemeris: $$ {\rm JD} = 2455682.4726 + 2.32517 E. $$

In shape, the light curve of KIC\,6864569 (see Fig.~6a) is remarkably similar to the light curve of KIC\,6278403, but the amplitude and period are different.
 For KIC\,6864569, the amplitude \mbox{$\Delta m = 0\fm03$}, and the period of brightness variability $P = 2\fd32517$ which is about twice as much as that for KIC\,6278403.

As well as for KIC\,6278403, the magnetic field strength of KIC\,6864569 does not exceed several hundred gauss and varies slightly with rotation (Fig.~6b). A small number of observation points does not allow us to confidently characterize the magnetic curve. However, the obtained criterion $\chi^2/n = 16.45$ supports the reliable field detection.

%\textbf{%}
Approximation of the observed spectrum of the star (Fig.~6c) gives
the following parameters: \linebreak \mbox{$T_{\rm
eff}\!=\!10\,612$\,K}, $\log g\!=\!4.03$,   \mbox{$v_e \sin i\!=\!37.02$\,km\,s$^{-1}$}, and $V_{R}=-17.30$~km s$^{-1}$ .

\subsection{KIC\,8161798 = BD\,+43 3223}

The best ephemeris: $${\rm JD} = 2455680.9554 + 2.20298 E. $$

Figure~7a shows the phase-convolved light curve of KIC\,8161798. It is a distinct double wave with primary and secondary minima almost coinciding in magnitude. The primary maximum is followed by the first minimum in the phase \mbox{$\phi = 0.25$}, then, the secondary maximum in the phase $\phi = 0.5$ followed by another minimum in the phase $\phi = $0.73. The brightness amplitude is relatively high: $\Delta m = 0\fm124$.

\begin{figure}[t] \includegraphics[width=0.99\linewidth, bb = 0 0 445 220,clip]{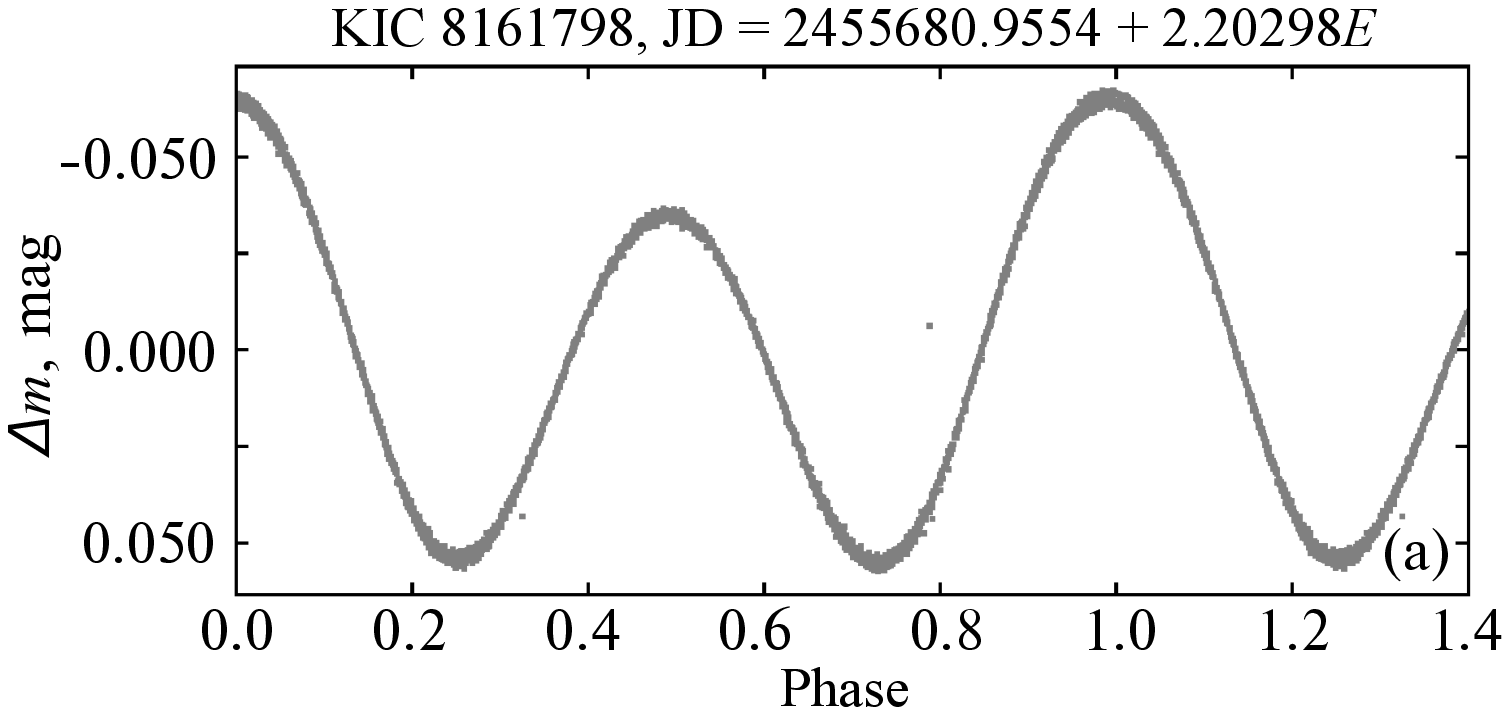} \\ \includegraphics[width=0.96\linewidth, bb = 0 0 590 375,clip ]{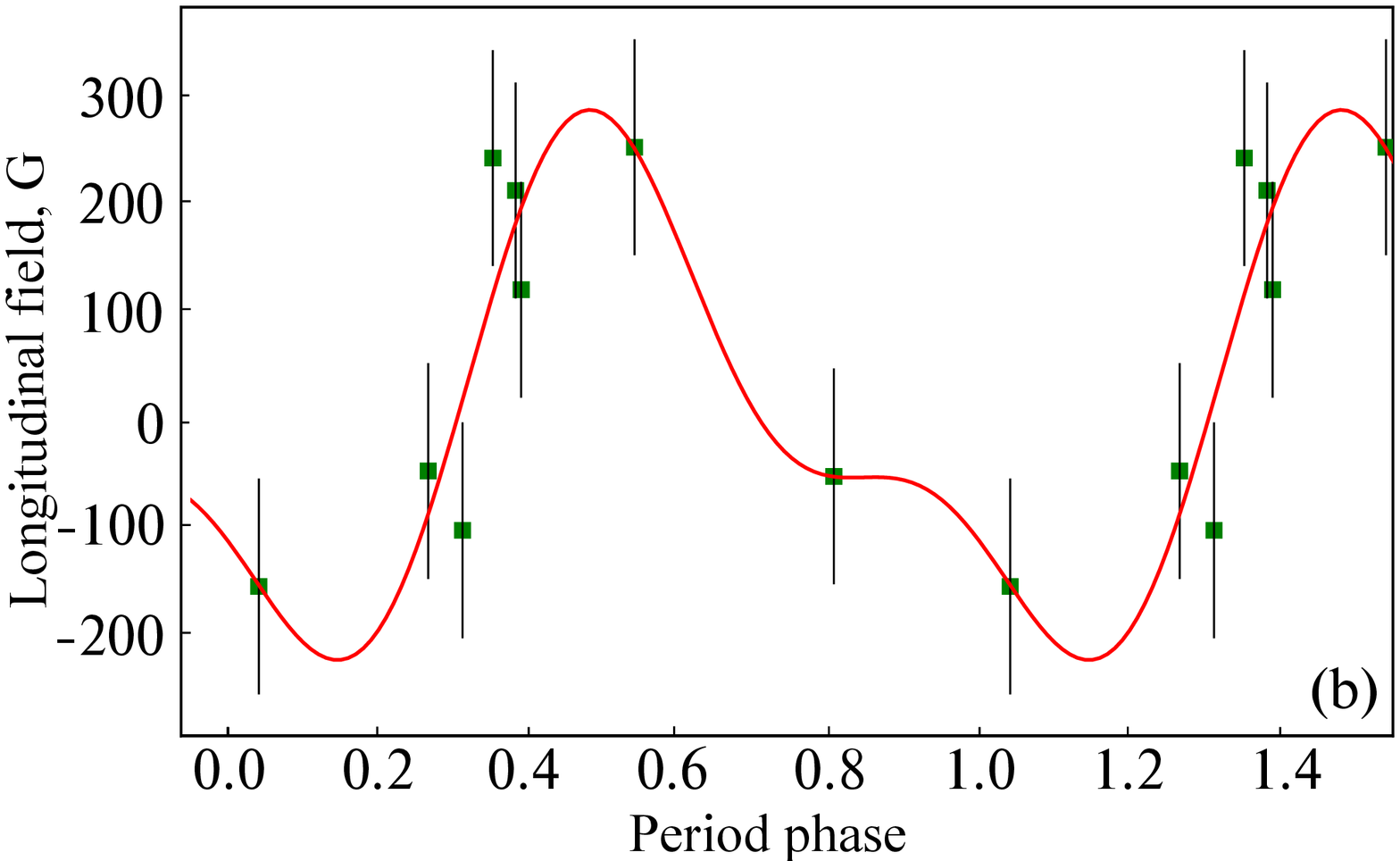}\\ \includegraphics[width=0.94\linewidth, bb = 0 0 620 410,clip]{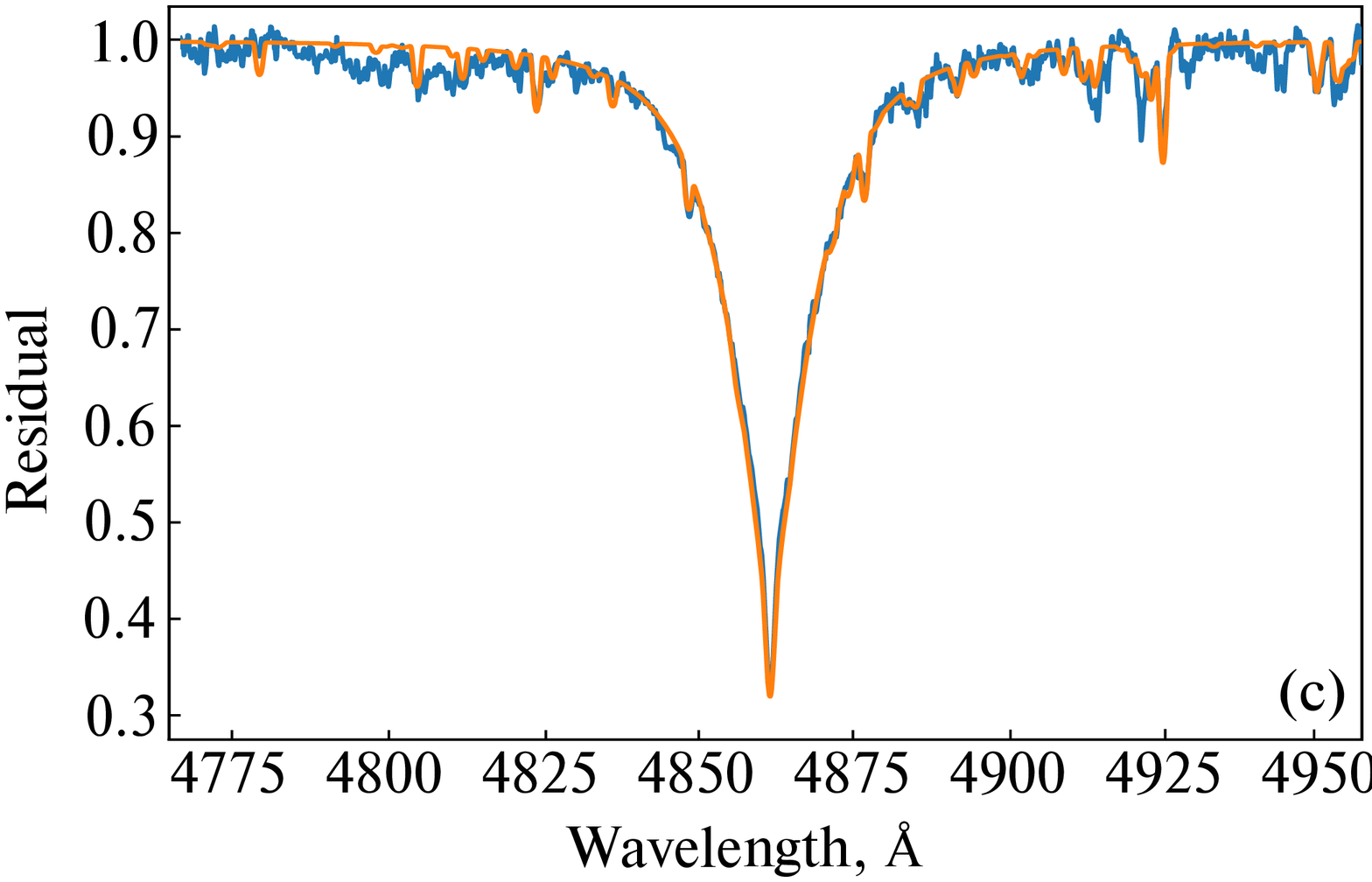} \caption{Same as in Fig.~1, for KIC\,8161798.} \label{fig7} \end{figure}

The criterion $\chi^2/n = 2.75$; the magnetic field is not reliably detected, but its variability is observed. The positive maximum of the magnetic field corresponds in phase to the secondary brightness maximum, the negative maximum is slightly shifted relative to the first brightness minimum.

%\textbf{%}
The parameters obtained on the basis of the spectrum approximation: $T_{\rm eff}=11\,664$~K, $\log g=4.00$, $v_e \sin i = 49.37$~km s$^{-1}$, and $V_{R}=-37.3$~km s$^{-1}$ (see Fig.~7c).

\subsection{KIC\,8324268 = HD\,189160}

The best ephemeris according to the TESS satellite: $${\rm JD} = 2458734.7007 + 2.00912 E.$$

 Figure~8 presents the resulting light curve of KIC\,8324268, the phase curve of the magnetic field, and the H${\beta}$-line modeling result in the spectrum of the object. The shape of the light curve is a single wave and almost a harmonic sine curve. Note its similarity with the light curve of KIC\,6065699. The brightening interval is slightly longer than the fading interval. The brightness maximum and minimum correspond to the phases: $\phi = 0.0$ and $\phi = 0.45$. The brightness amplitude \mbox{$\Delta m = 0\fm026$}.

\begin{figure} \includegraphics[width=0.99\linewidth, bb = 0 0 445 220,clip]{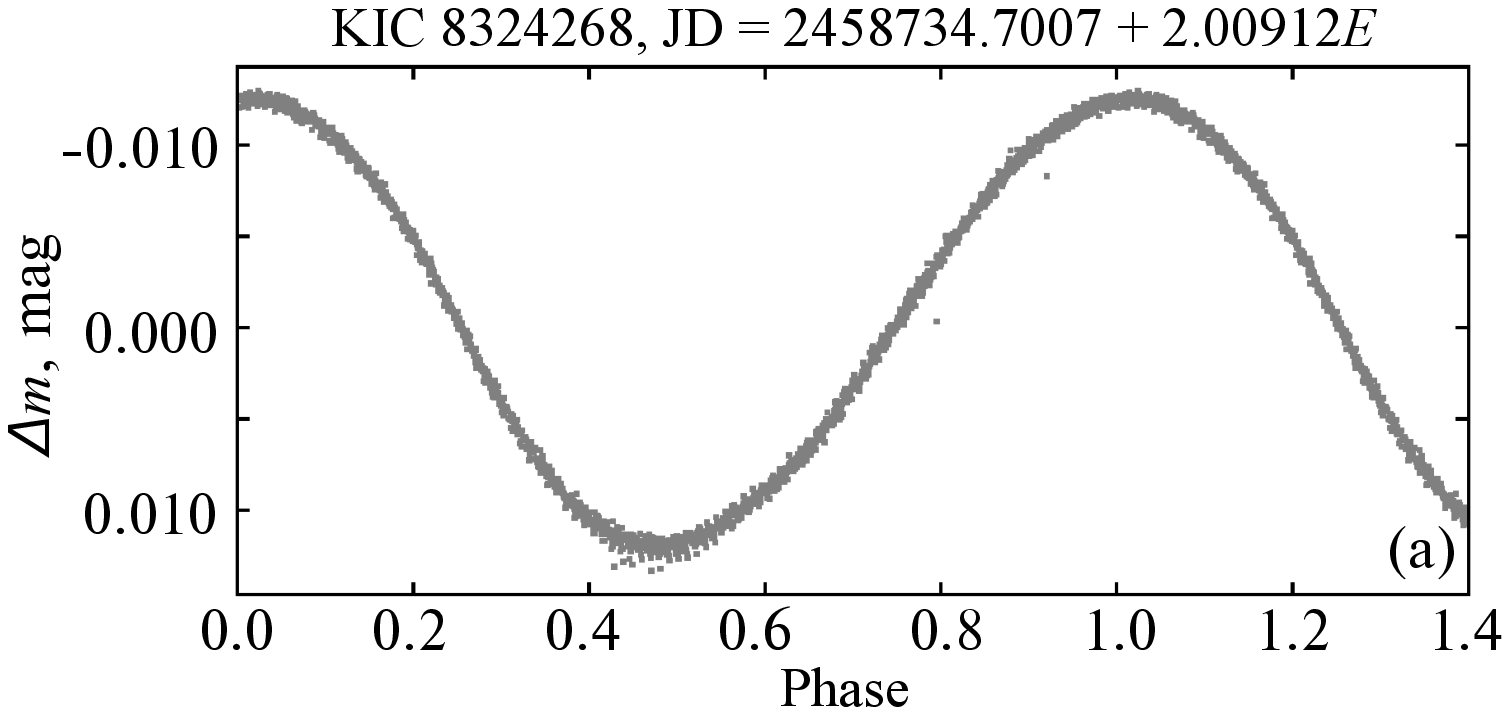} \\ \includegraphics[width=0.96\linewidth, bb = 0 0 590 375,clip ]{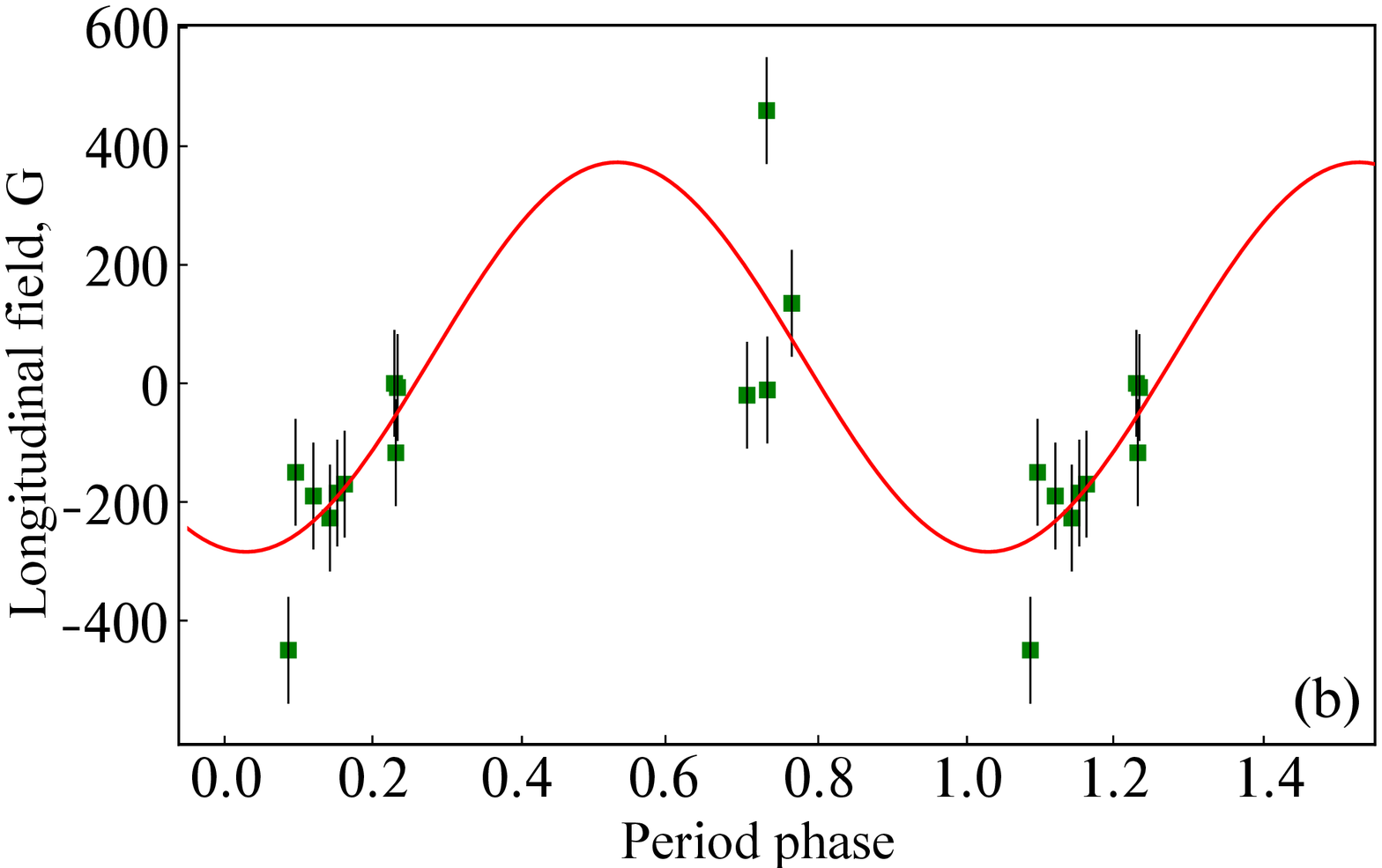}\\ \includegraphics[width=0.94\linewidth, bb = 0 0 620 410,clip]{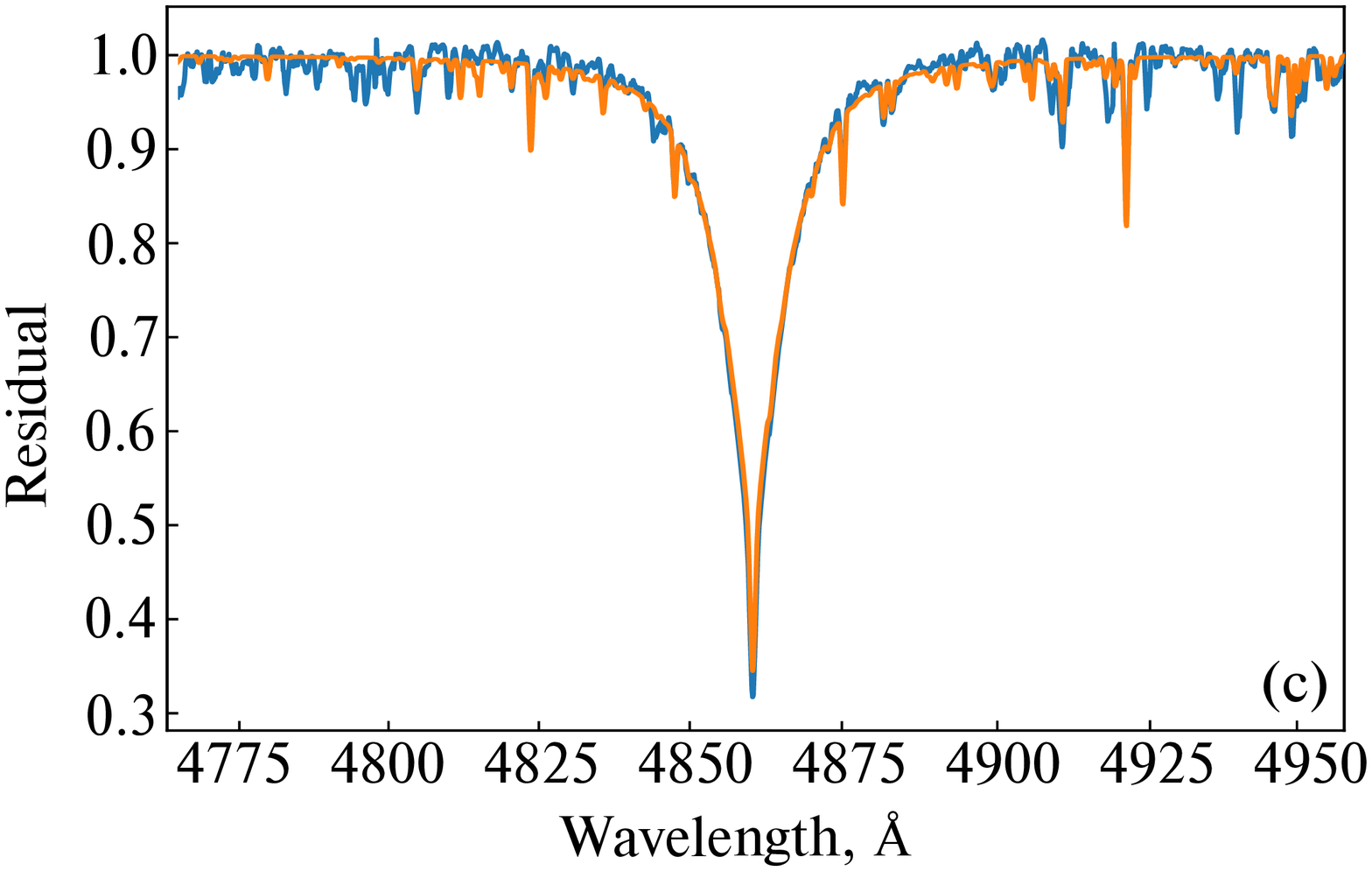} \caption{Same as in Fig.~1, for KIC\,8324268.} \label{fig8} \end{figure}

Magnetic field measurements show large scatter (see Fig.~8b), which makes it difficult to approximate by a single wave, but the criterion $\chi^2/n = 5.88$ indicates its reliable detection.

%\textbf{%}
The spectrum approximation (Fig.~8c) gives the following parameters: $T_{\rm eff}=12\,996$~K, \mbox{$\log g=3.85$}, $v_e \sin i = 26.0$~km s$^{-1}$, and \mbox{$V_{R}=-17.5$~km s$^{-1}$}.

\subsection{KIC\,10324412 = HD\,176436} When analyzing the time
series from the TESS satellite, the ephemeris was obtained: $$ JD
= 2455680.7844 + 1.73150 E. $$

The light curve of the object is a
double wave with a primary and secondary maximum near the phases
$\phi = 0.0$ and $\phi = 0.6$ respectively. The brightness
minimum is near the phase $\phi = 0.3$, the brightness variation
amplitude $\Delta m = 0\fm03$. (see Fig.~9a).

The magnetic field is very weak, almost at the detection limit showing a single sinusoid with the minimum near the primary maximum of the light curve and with the maximum near the secondary (Fig.~9b). The estimate $\chi^2/n = 0.89$ does not allow one to assert the reliable detection of a magnetic field.

The parameters determined during the spectrum approximation: $T_{\rm eff}=10\,142$~K, \mbox{$\log g=3.99$}, $v_e \sin i = 77.73$~km s$^{-1}$, $V_{R}=1.39$~km s$^{-1}$, and ${\rm [{\rm M/H}]} = 0.544$. The observed and synthetic spectra are shown in Fig.~9c.

\begin{figure} 
\includegraphics[width=0.99\linewidth, bb = 0 0 445 220,clip]{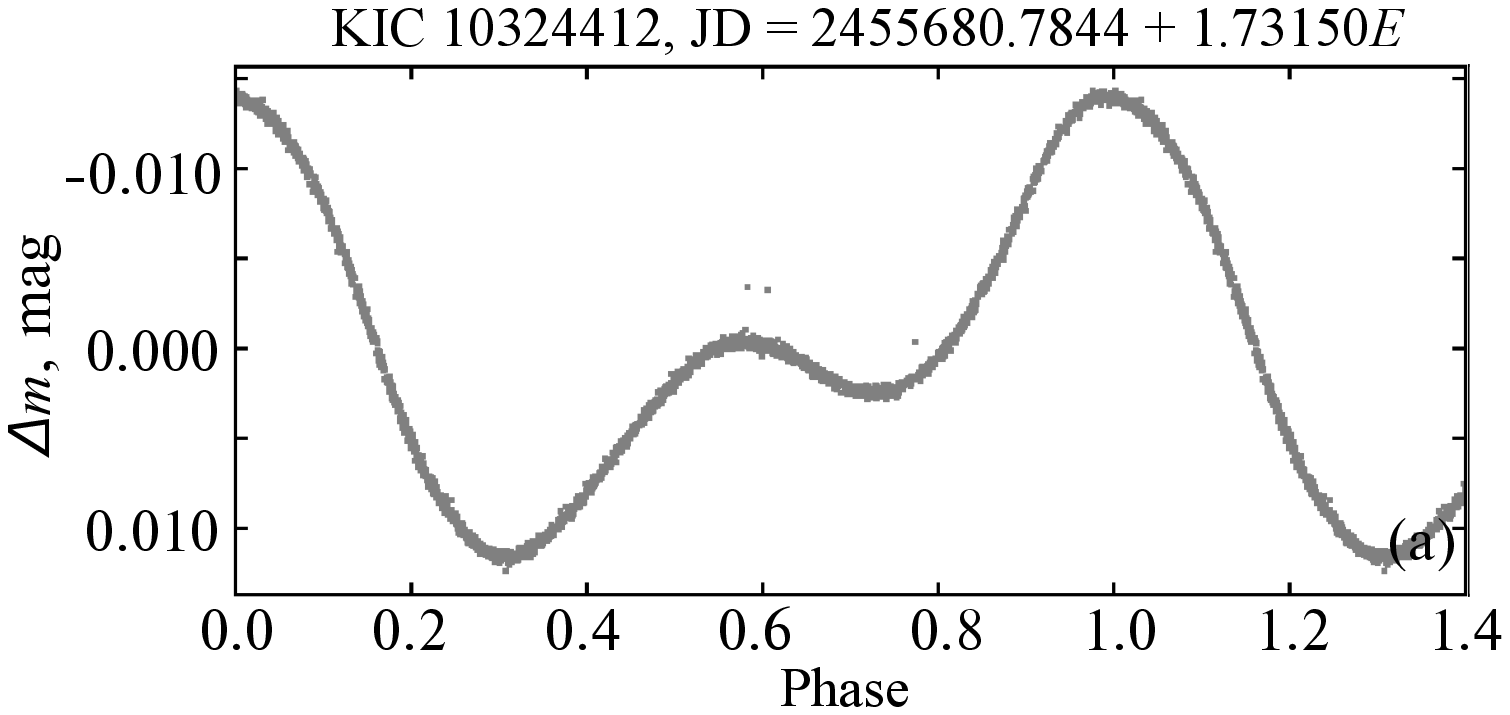} \\
 \includegraphics[width=0.96\linewidth, bb = 0 0 590 375,clip ]{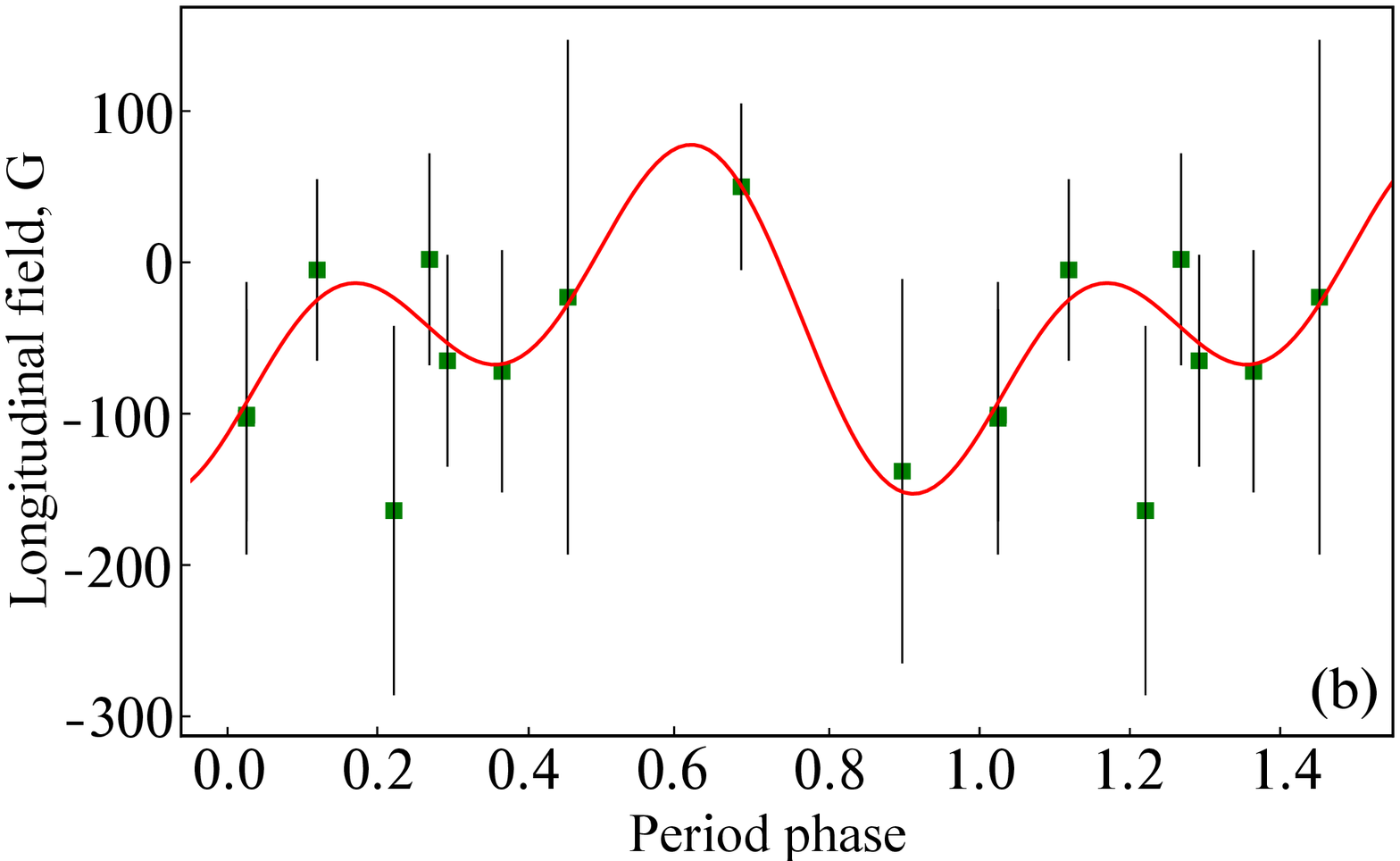}\\
  \includegraphics[width=0.94\linewidth, bb = 0 0 620 410,clip]{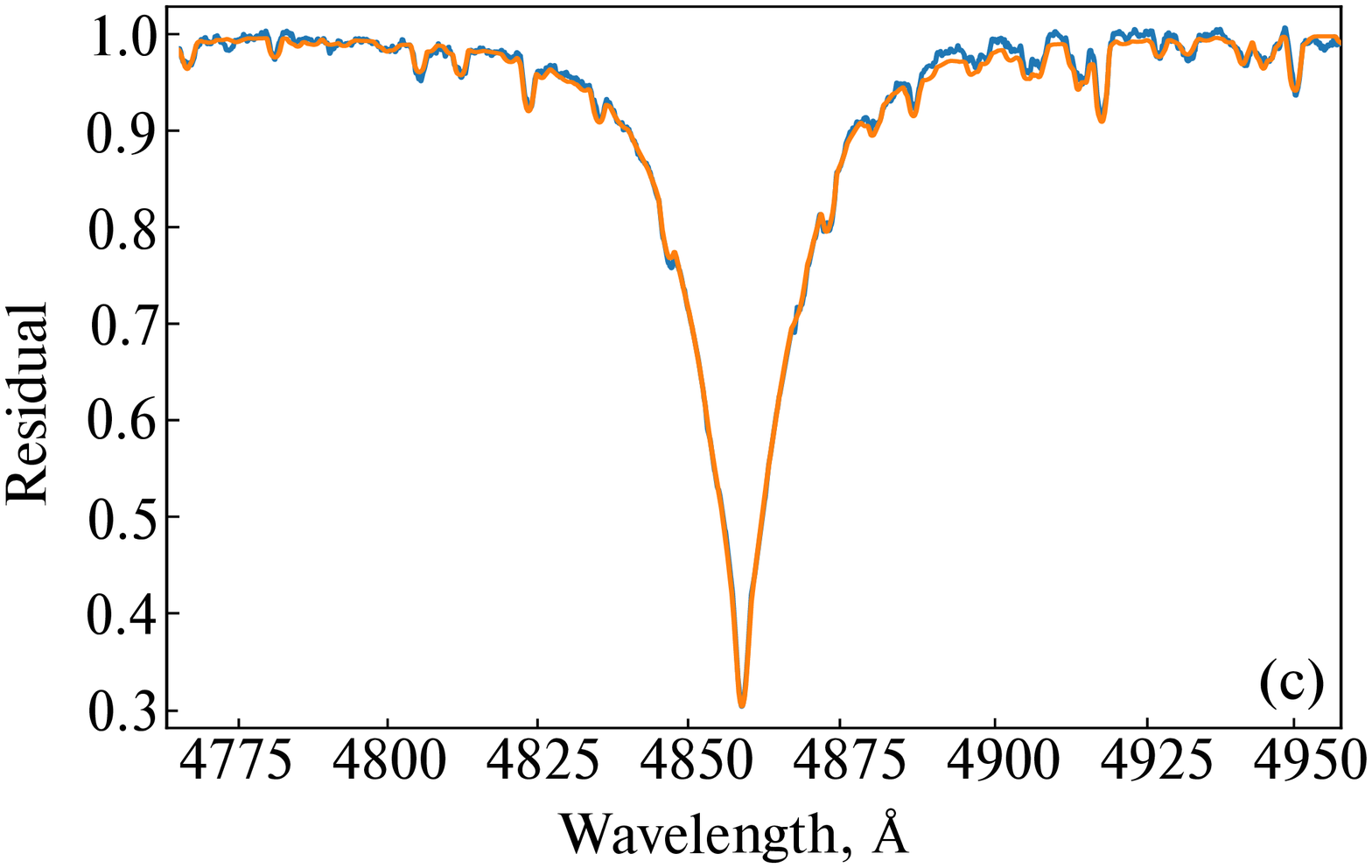}
\caption{Same as in Fig.~1, for KIC\,10324412.} \label{fig9} \end{figure}

\subsection{KIC\,11560273 = HD\,184007}

\begin{figure}  \includegraphics[width=0.99\linewidth, bb = 0 0 445 220,clip]{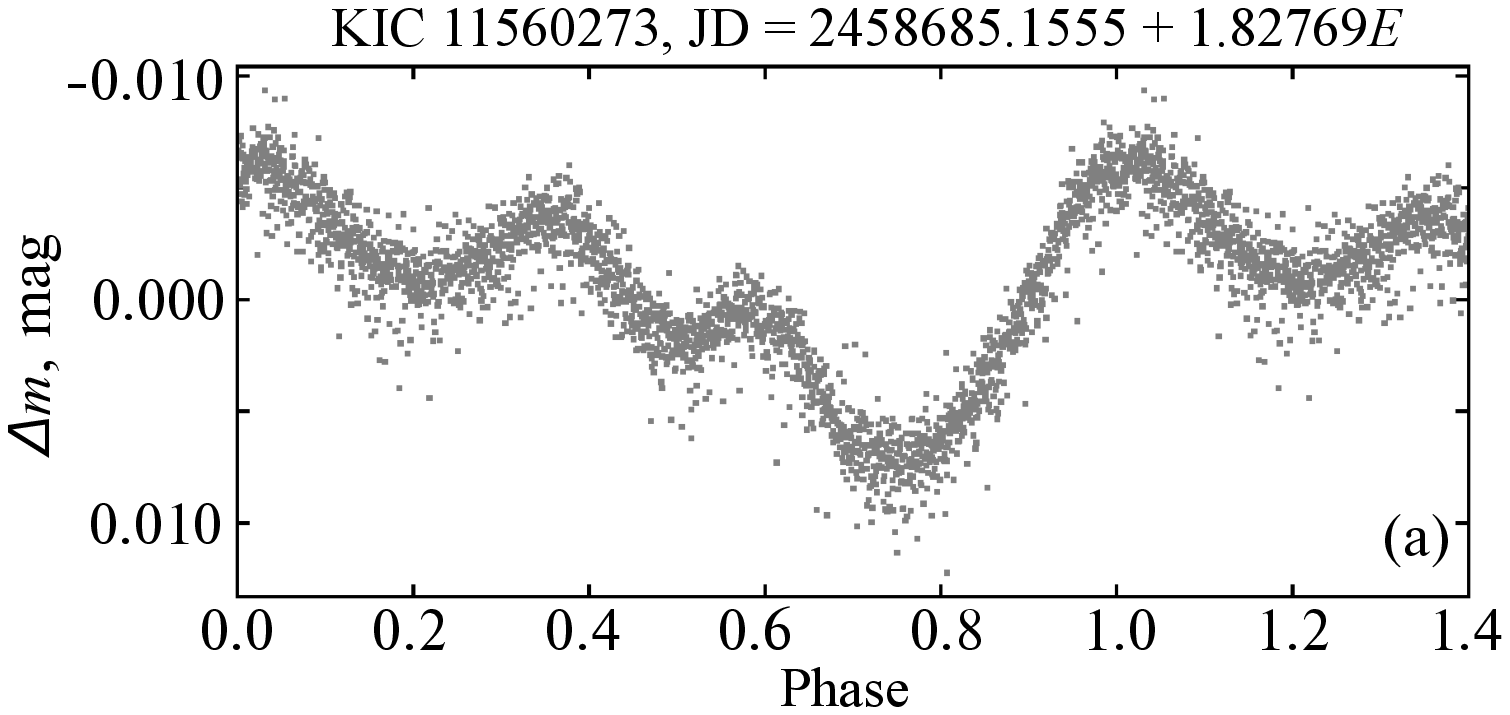} \\ \includegraphics[width=0.96\linewidth, bb = 0 0 590 375,clip]{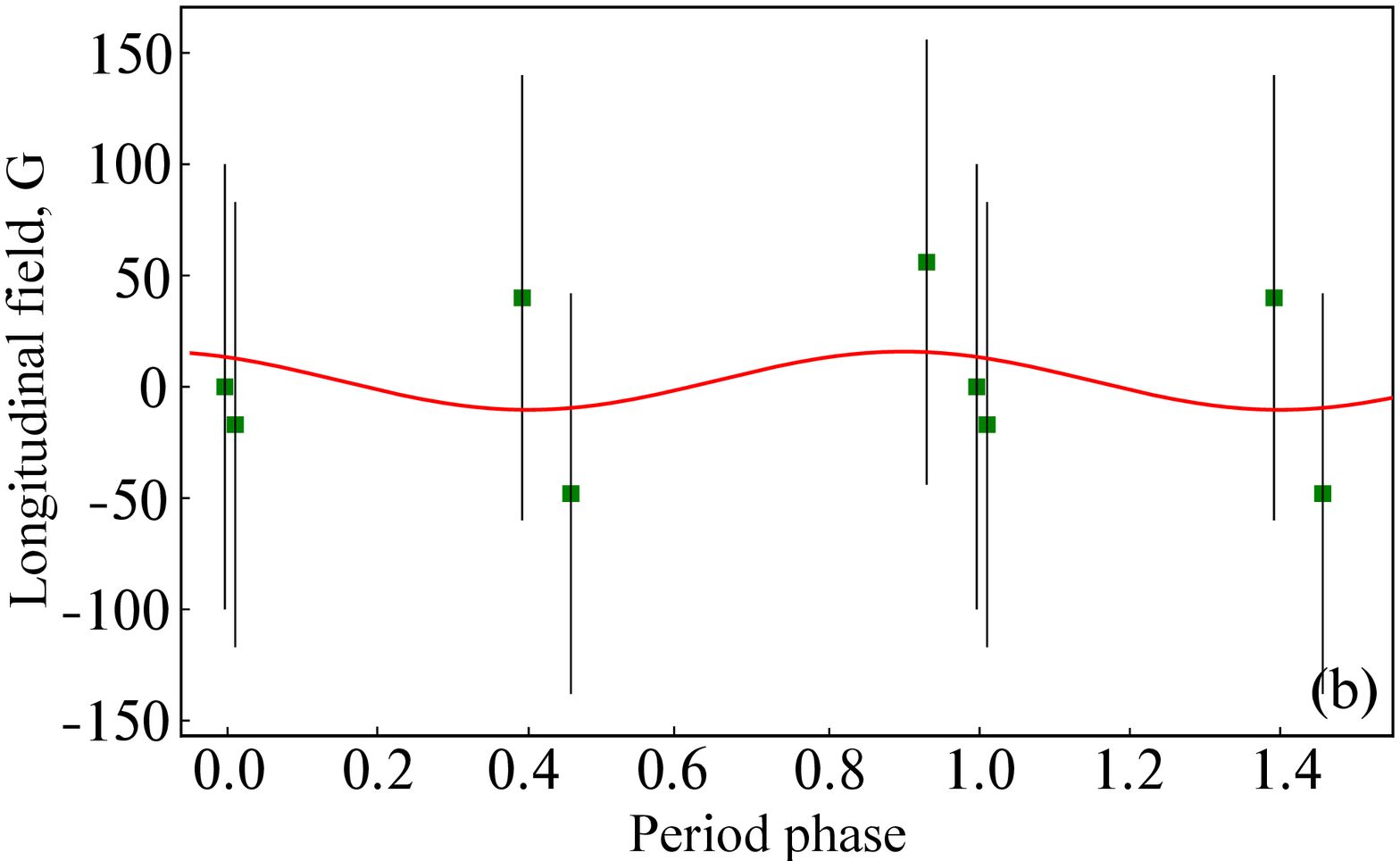}\\ \includegraphics[width=0.94\linewidth, bb = 0 0 620 410,clip]{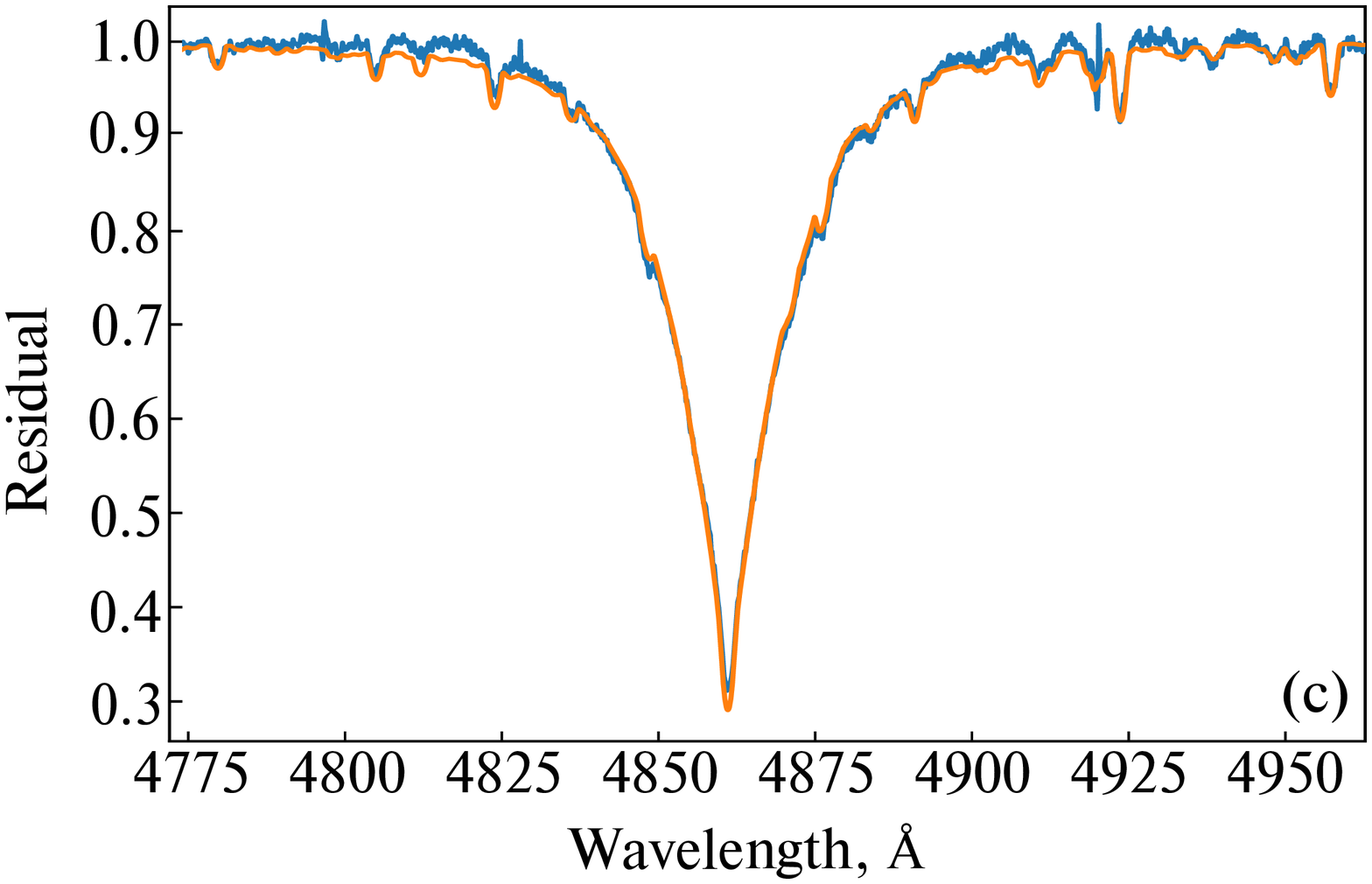} \caption{Same as in Fig.~1, for KIC\,11560273.} \label{fig10} \end{figure}

The best ephemeris: $$ {\rm JD}=2458685.1555 + 1.82769 E. $$

Figure~10a shows the phased light curve of the object. Its shape is another example of the previously noted diversity of light curves of mCP star candidates. The total amplitude of the brightness variation \mbox{$\Delta m = 0\fm02$}.

Measurements of the longitudinal magnetic field strength are burdened with large errors (Fig.~10b). The estimate $\chi^2/n = 0.16$ indicates that no magnetic field is detected.

Figure~10c presents the observed spectrum of the object and the result of its synthetic approximation with the following model parameters: \mbox{$T_{\rm eff}=9\,744$~K}, $\log g=3.77$, \mbox{$v_e \sin i = 73.8$~km s$^{-1}$,} and \linebreak \mbox{$V_{R}=-15.5$}~km s$^{-1}$.

\section{CONCLUSION}

We present the first results of studying the variability of a part
of the mCP star candidate sample that we have compiled using the
analysis of the Kepler satellite photometric data. The proposed
candidate selection method demonstrates its effectiveness: the
spectra of all the stars in the sample contain peculiar lines
which has been previously demonstrated in the paper by
{H{\"u}mmerich et~al. (2018), observations for which have been
obtained with the Zeiss-1000 SAO RAS telescope. The magnetic
fields are reliably detected in six stars out of the ten objects
selected for monitoring: KIC\,4180396, KIC\,5264818, KIC\,5473826,
KIC\,6065699, KIC\,6864569, and \linebreak \mbox{KIC\,8324268}. For the remaining
four objects, the field is detected at the level of instrumental
measurement errors during observations with the MSS at the SAO RAS
BTA.

Curiously, the shapes of the light curves of the objects show
extraordinary variety. Most likely, this can be explained by the
spotted surface structures characteristic of CP stars. For the
future, there are plans to study the effect of spots of individual
chemical elements on the shape of the photometric light curves.

At the present stage, it is difficult to talk about regularities
in correlation between the behavior of the magnetic field and
photometric curves. Only for three objects (KIC\,4180396,
KIC\,5264818, and \mbox{KIC\,8161798}) positive or negative extrema of
the magnetic curves correspond to the photometric extrema (primary
or secondary). To eliminate the random factor, a large sample of
objects is needed.

%\textbf{%} %
%\textbf{
Spectropolarimetric monitoring of the photometric candidate sample will be continued.

%}
\begin{acknowledgments} The authors are grateful to D.~O. Kudryavtsev and E.~G. Sendzikas for their help with the observations.

The authors are grateful to the anonymous reviewers for valuable comments which made it possible to improve the paper contents and clarify some critical points.

This paper contains the data collected by the TESS mission from the MAST data archive of the Space Telescope Science Institute (STScI). Funding for the TESS mission is provided by the NASA Explorer program. STScI is administered by the Association of Universities for Research in Astronomy under the NASA contract NAS 5--26555.

The paper uses the VALD database operating at the Uppsala University, the Institute of Astronomy of the Russian Academy of Sciences in Moscow, and the University of Vienna was used.

Observations with the SAO RAS telescopes are supported by the Ministry of Science and Higher Education of the Russian Federation. Upgrading of the instruments is carried out within the framework of the ``Science and Universities'' national project.

%This paper includes data collected with the TESS mission, obtained from the MAST data archive at the Space Telescope Science Institute (STScI). Funding for the TESS mission is provided by the NASA Explorer Program. STScI is operated by the Association of Universities for Research in Astronomy, Inc., under NASA contract NAS 5--26555.
%This work has made use of the VALD database, operated at Uppsala University, the Institute of Astronomy RAS in Moscow, and the University of Vienna.
\end{acknowledgments}

\section*{FUNDING} The observation part of the study and data
reduction (IY) were carried out with the financial support of the
RFBR within the framework of the scientific project
No.~19-32-60007.

Analyzing and building the magnetic field phase curves,
determination of physical parameters (EAS, IIR, and AVM) were
carried out with partial financial support from the Russian
Science Foundation (RSF) No.~21-12-00147.

\section*{CONFLICT OF INTEREST} The authors declare no conflict of interest regarding the publication of this paper.

%\bibliographystyle{AstroBull}
%\bibliography{references}
 \end{document}